\documentclass[a4paper,12pt]{article}
\usepackage{mathptmx}
\usepackage{epsfig}
\usepackage{float}
\usepackage{amssymb}
\usepackage{latexsym}
\usepackage{authblk}

\usepackage{amsfonts}
\newcommand{\beq}{\begin{equation}}
\newcommand{\eeq}{\end{equation}}
\newcommand{\beqa}{\begin{eqnarray}}
\newcommand{\eeqa}{\end{eqnarray}}

\usepackage{natbib}
\usepackage{graphicx}
\usepackage{multirow}
\begin{document}



\title{Analysis of the ground level enhancements on 14 July 2000 and on 13 December 2006 using neutron monitor data}
\author[1]{A.L. Mishev}
\author[1,2]{I.G. Usoskin}
\affil[1]{ReSoLVE Center of Excellence, University of Oulu, Finland.}
\affil[1]{Sodankyl\"a Geophysical Observatory (Oulu unit), University of Oulu, Finland.}
\maketitle
\begin{abstract}

On the basis of neutron monitor data we estimate the energy spectrum, anisotropy axis direction and pitch-angle distribution of solar energetic particles during two major ground level enhancements (GLE 59 on 14 July 2000 and GLE 70 on 13 December 2006). For the analysis we use a newly computed neutron monitor yield function. The method consists of several consecutive steps: definition of the asymptotic viewing cones of neutron monitor stations considered for the data analysis by computations of cosmic ray particles propagation in a model magnetosphere with the MAGNETOCOSMICS code; computation of the neutron monitor model responses and derivation of the solar energetic particle characteristics on the basis of inverse problem solution. The pitch-angle distribution and rigidity spectrum of high-energy protons are obtained as function of time in the course of ground level enhancements. A comparison with previously reported results is performed and reasonable agreement is achieved. A discussion of the obtained results is included as well their possible application is included.

\end{abstract}

 \normalsize

\label{cor}{\small For contact: alexander.mishev@oulu.fi; ilya.usoskin@oulu.fi}


\section{Introduction}

A thorough analysis of solar energetic particle (SEP) events provides crucial information on particle scattering and transport
in the interplanetary medium as well as understanding of their acceleration mechanisms \citep{Deb88, Lockwood1990, Kallenrode1992, Reames99}. Some solar flares and eruptive events, such as coronal mass ejections (CMEs), can accelerate protons and other ions to high energies \citep[][and references therein]{Cliver04, Dor06, Reames2009a, Reames2009b, Aschwanden12}. Such SEPs can enter the Earth atmosphere sporadically, with a greater probability during maximum and declining phase of a solar activity cycle \citep[\textit{e.g.}][]{Shea90}.

High-energy particles, mostly protons and $\alpha$-particles, of extra-solar origin known as galactic cosmic rays (GCR) constantly hit the Earth atmosphere. They induce a complicated atmospheric nuclear--electromagnetic--muon cascade (see \citet{Dorman04, Bazilevskaya08} and references therein). Occasionally the energy of SEPs is large enough ( $\ge$ 1 GeV/nucleon) to initiate a similar atmospheric cascade leading to an enhancement of count rate of ground-based detectors, specifically neutron monitors (NMs). This special class of SEP events is called ground-level enhancements (GLEs). The worldwide network of NMs \citep [\textit{e.g}.][]{Mav11} represents a multi-instrumental tool for continuous monitoring of the intensity of CR particles as well as the registration of GLEs \citep{Sim53, Hat71}. 

The NM stations are spread over the globe at multiple geographic locations in order to obtain a complete picture of cosmic rays in space, since their intensity is not uniform around the Earth \citep{Bieber95}. This is particularly important for GLEs, which possess an essential anisotropic part, specifically during the event's onset \citep{Deb88, Shea90, Vas06, But09}. Therefore, measurements performed by the worldwide network of NMs form the basis to assess the spectral and angular characteristics of SEPs near Earth. In addition, it was recently shown that spectral and angular characteristics of SEPs derived using NM data are model dependent \citep{But13a}, mainly due to assumed NM yield function(s), magnetospheric model(s) (for details see \citet{But13b}). In this study we use an established method for the analysis of GLEs based on NM data but applying a newly computed neutron monitor yield function \citep{Mishev14c}. We study two major events, namely GLE 59 on 14 July 2000 and GLE 70 on 13 December 2006.

\section{Modelling the neutron monitor response}
In order to derive the SEPs characteristics, it is necessary to establish a relationship between NM count rates and the primary particle flux, considering the full complexity of particle transport in the geomagnetosphere and in the Earth atmosphere \citep{Debrunner68, Hat71, Sma00, Dorman04, Desorgher2009, Mis13a}. A convenient formalism for relation between NM count rate and primary particle flux is based on yield function \citep[for details see \textit{e.g.}][and references therein]{Hat71, Cle00}. In this study, we use a recently computed NM yield function, which considers the finite lateral extent of cosmic ray induced atmospheric cascade and provides good agreement with experimental latitude surveys \citep{Mis13a, Mis13b}. 

An analysis of GLEs based on NM data consists of several consecutive steps: determination of asymptotic viewing cones of the NMs by computation of particle trajectories in a model magnetosphere; assumption of an initial guess of the inverse problem; application of an optimization procedure (inverse method) for derivation of the primary SEPs' energy spectrum, anisotropy axis direction and pitch angle distribution. A detailed description of the method is given elsewhere \citep{Mishev14c}. The method is similar to that used by \citet{Shea82, Humble91, Cramp97, Bom06, Vas06, Vas08}.

The relative count rate increase of a given NM can be expressed as:

\begin{equation}
\frac{\Delta N(P_{cut})}{N} =\frac{\int_{P_{cut}}^{P_{max}}J_{||sep}(P,t)Y(P)G (\alpha(P,t)) dP}{\int_{P_{cut}}^{\infty}J_{GCR}(P,t)Y(P)dP} ,
\label{simp_eqn1}
   \end{equation}
\noindent where $J_{||sep}$ is the rigidity spectrum of the primary solar particles in the direction of the maximal flux, $J _{GCR}(P,t)$ is the rigidity spectrum of GCR at given time $t$ with the corresponding modulation, Y(P) is the NM yield function, $G(\alpha(P,t))$ is the pitch angle distribution of SEPs, $N$ is the count rate due to GCR, $\Delta N(P_{cut})$ is the count rate increase due to solar particles, $P_{cut}$ is the minimum rigidity cut-off of the station, accordingly $P_{max}$ is the maximum rigidity of SEPs considered in the model, assumed to be 20 GV, which is sufficiently high for SEPs. The fractional increase of the count rate of a NM station represents the ratio between the NM count rates due to SEPs and GCR averaged over 2 hours before the event's onset.  In our model we assume a modified power law rigidity spectrum of SEP similarly to \citet{Cramp95, Cramp97, Bom06, Vas08}:

\begin{equation}
    J_{||}(P)= J_{0}P^{-(\gamma+\delta\gamma(P-1))} ,
        \label{simp_eqn2}
   \end{equation}

\noindent  where $J_{||}$ is particle flux arriving from the Sun along the axis of symmetry whose direction is defined by geographic coordinate angles $\Psi$ and $\Lambda$, $\gamma$ is the power-law spectral exponent at rigidity P = 1 GV,  $\delta\gamma$ is the rate of the spectrum steepening. The pitch angle distribution is assumed to be a Gaussian:

\begin{equation}
    G(\alpha(P)) \sim exp(-\alpha^{2}/\sigma^{2}) , 
        \label{simp_eqn3}
   \end{equation}

\noindent where $\alpha$ is the pitch angle, $\sigma$ is parameter corresponding to the width of the pitch angle distribution. The pitch angle is defined as the angle between the asymptotic direction and the axis of anisotropy. Therefore according to equations (1) to (3) six parameters have to be determined: $J_{0}$, $\gamma$ , $\delta\gamma$,  $\Psi$ and $\Lambda$, $\sigma$. 

In our study, for the computation of rigidity cut-offs and asymptotic directions of the allowed trajectories \citep{Cook91} we use the combination of International Geomagnetic Reference Field (IGRF) geomagnetic model (epoch 2010) as the internal field model \citep{Lan87} and the Tsyganenko 89 model as external field \citep{Tsy89}. This combination provides the most efficient computation of asymptotic directions as well as a  balance between simplicity and realism \citep{Kud04, Kud08, Nev13}.  All computations of particle's transport in the geomagnetic field are performed with the MAGNETOCOSMICS code \citep{Des05}. 

For the GCR spectrum we apply a parametrisation based on the force-field model \citep{Gle68, Cab04}. Full details of the application of the model are given elsewhere \citep{Uso05}. The solar modulation parameter is taken from \citet{Uso11a}. For the solution of the inverse problem we use the Levenberg-Marquardt algorithm (LMA) \citep{Lev44, Mar63} with the Minpack code \citep{Mor80}. The optimization is performed by minimization of the difference between the modelled NM responses and the measured NM responses \textit{i.e.} optimization of the functional $\mathcal{F}$ over the vector of unknowns and $m$ NM stations:

\begin{equation}
\mathcal{F}=\sum_{i=1}^{m} \left[\left(\frac{\Delta N_{i}}{N_{i}}\right)_{mod.}-\left(\frac{\Delta N_{i}}{N_{i}}\right)_{meas.}\right]^{2} ,
\label{simp_eqn4}
   \end{equation}

\section{Results of the analysis}
In total 16 GLEs were observed during the Solar Cycle 23 \citep{Andriopoulou2011a, Andriopoulou2011b, Gopalswamy2012}. The strongest event, namely GLE 69 on 20 January 2005 was analysed elsewhere \citep{Vas06b, Plainaki2007, Bombardier2008, Perez-Peraza2008, But09, Bieber2013}. Here we focus on two major events: GLE 59 on 14 July 2000 also known as the Bastille day event and GLE 70 on 13 December 2006.

\subsection{The Bastille day GLE 59 on 14 July 2000}
The second decade of July 2000  was characterized by intense solar activity extending from 10 to 15 of July. During this period three
X-class flares (including the Bastille Day flare) and two halo CMEs were produced \citep{Dryer2001}. The GLE 59 event was related to the Bastille day X5.8/3B solar flare and the associated full halo
CME. It started at 10:03 UT, reached peak at 10:24 UT and ended at 10:43 UT \citep{Klein2001a}. Accordingly, the GLE onset began between 10:30 and 10:35 UT at several
stations (Figure 1). The strongest NM increases were observed at the South Pole (58.3 $\%$) and SANAE (54.4 $\%$) compared to pre-increase levels. In general the event was characterized by a large anisotropy in its initial phase \citep{Bieber02, Bom06, Vas06}.

\begin{figure}[H]
   \centering
   \includegraphics[width=\textwidth]{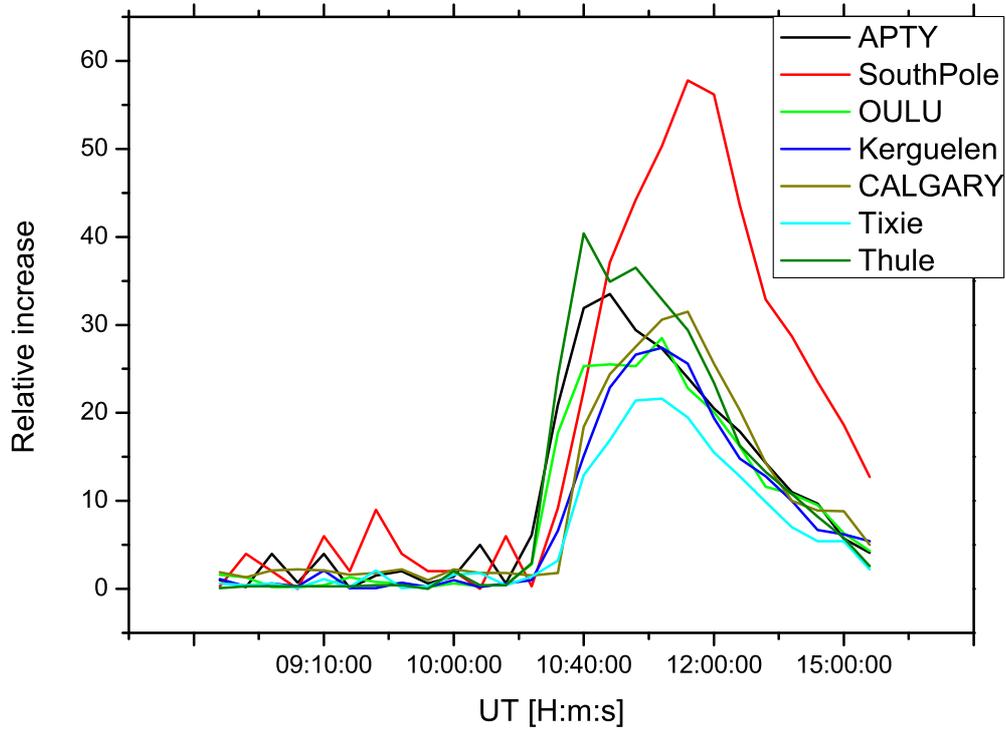}
      \caption{Profile of the time variation of Apatity, Calgary, Kerguelen, South Pole, Oulu, Thule and Tixie NMs relative increase during GLE 59 on 14 July 2000. 
              }
         \label{Fig1}
   \end{figure}

An illustration of several computed asymptotic cones used for our analysis is shown in Figure 2. The full list of NMs used in this analysis is given below (Table 3). The computations were carried out with the MAGNETOCOSMICS code using Tsyganenko 1989 (external field model) and International Geomagnetic Reference Field (IGRF) (internal field model) geomagnetic models, adjusted to the measured $K_{p}$ index and 2000 epoch. Here we present the asymptotic directions in the rigidity range  from 1 to 5 GV  in order to demonstrate the range of maximal response, while in the analysis we used the 1-20 GV rigidity range.

\begin{figure}[H]
   \centering
   \includegraphics[width=\textwidth]{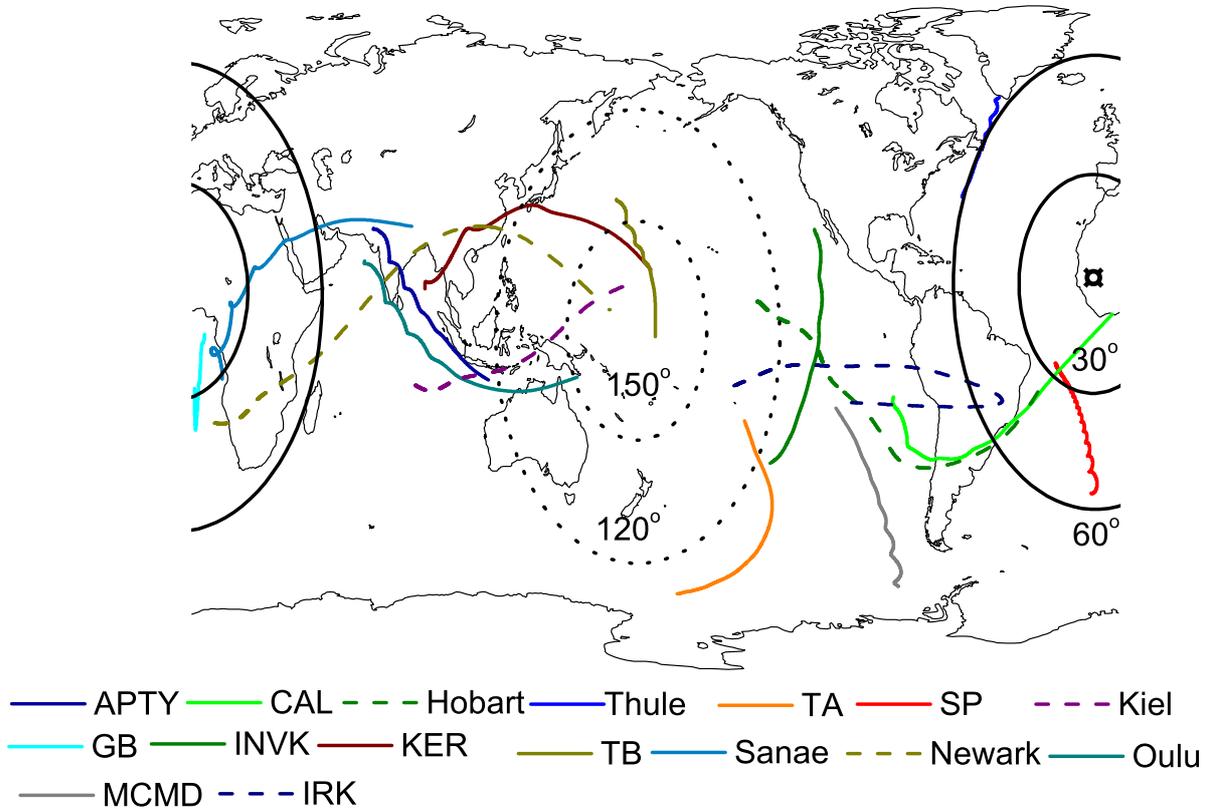}
      \caption{Calculated NM asymptotic directions during GLE 59 on 14 July 2000 at 10:30 UT. The small oval represents the derived apparent source position during the event onset. The lines of equal pitch angles relative to the derived anisotropy axis are plotted for 30$^{\circ}$, 60$^{\circ}$, 150$^{\circ}$ and 120$^{\circ}$. The asymptotic directions of polar NMs are plotted with solid lines, while mid-latitude NMs are plotted with dashed lines.
              }
         \label{Fig2}
   \end{figure}

For the analysis we considered 5-min NM data retrieved from GLE database \citep{Usoskin2015a}. The derived rigidity spectra with the anisotropy characteristics of the high-energy SEP are presented in Figure 3. During the event's onset, SEPs had a hard rigidity spectrum and strong anisotropy as it was observed by NM stations with small pitch-angles. 

The left-hand panel of Figure 3 demonstrates the obtained rigidity spectrum during various stages of the event. The corresponding pitch angle distributions are presented in the right-hand panel of Figure 3. One can see that the rigidity spectrum was gradually softening throughout the event. The steepening $\delta\gamma$ varied throughout the event. It resulted in a moderate steepening of the spectrum during the early phase and only a slight steepening during the late phase of the event. Moreover, $\delta\gamma$ vanished after 13:00 UT so that the derived rigidity spectra can be described  by a pure power law. The derived anisotropy was high during the event's onset. It rapidly dropped to a low, relatively steady level so that the SEPs angular distribution considerably broadened out after the initial phase of the event. We can conclude that the event was isotropised very fast, leading to a nearly isotropic SEP flux after 11:00 UT. The derived spectral and anisotropy characteristics are summarized in Table 1. 

The quality of the modelling is demonstrated by comparison between the modelled and observed responses of NMs during the GLE 59. Here we present a comparison for several NM stations (Figure 4), but the quality of the fit is similar for the other NM stations. Since there is no dramatic change of the derived SEP characteristics after 11:00 UT (gradual softening of the rigidity spectrum and isotropic phase), in order to present detailed information throughout the whole event, the X axis (Time) in Figure 4 is not uniform.

A detailed analysis of confidence limits of the derived model parameters is rather difficult, because of strong non-linearity and complexity of the model as well as a significant correlation between the fit parameters \citep{Den83, Aster2005}. We examine the squared difference between the observed and calculated increases. We achieve less than 5 $\%$ relative difference between modelled and observed NM relative increases. Note that in the analysis we employed a natural initial guess similarly to \citet{Bom06}, \textit{i.e.} that the apparent source position is located along the interplanetary magnetic field (IMF) line derived from the ACE satellite measurements (otherwise the initial guess is similar to that of \citet{Cramp95}). According to our analysis a small variation of the input (initial guess) does not alter the solution. In case when the initial guess is far from the local minimum (the LMA converges to the global minimum only if the initial guess is close to the final solution \citep{Den83}), the derived solution of the inverse problem possess greater residual compared to that with a natural initial guess or it is not physical.

\begin{figure}[H]
   \centering
   \includegraphics[width=\textwidth]{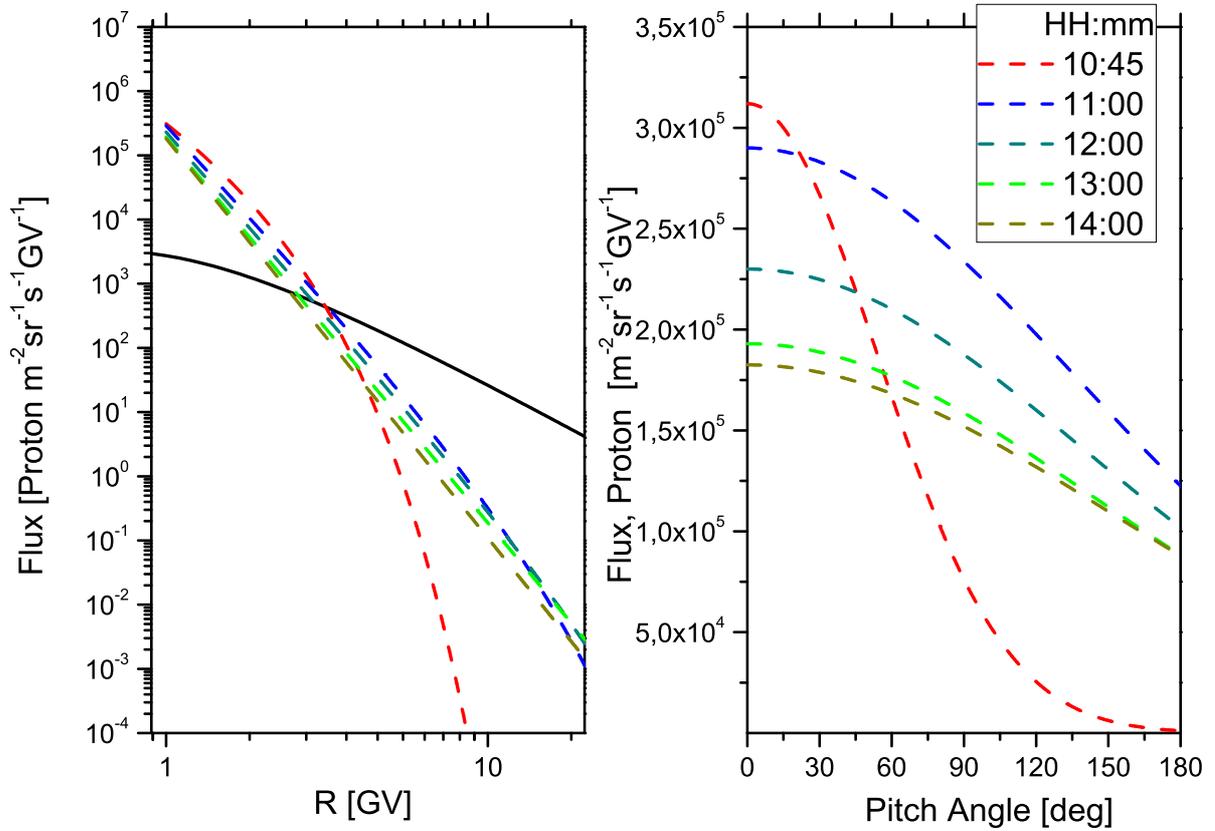}
      \caption{Derived rigidity spectra and pitch angle distributions of SEPs during the GLE 59. The SEP flux $J_{||}$ is according to Equation (2) \textit{i.e.} flux arriving from the Sun along the axis of symmetry. Time (UT) refers to the end of the corresponding five minute interval. The solid line in the left panel denote GCR flux.
              }
         \label{Fig3}
   \end{figure}

\begin{table*}[htp]
\caption{Derived spectral and angular characteristics for GLE 59 on 14 July 2000}             
\label{table:A1}      
\centering          
\begin{tabular}{c c c c c c c }     
    

\hline       
Integration interval [UT] & $J_{0}$ [$m^{-2} s^{-1} sr^{-1} GV^{-1}$] & $\gamma$ & $\delta\gamma$ & $\sigma^{2}$ [rad $^{2}$] & $\Psi$ [degrees]& $\Lambda$ [degrees] \\ 
\hline                    
10:35--10:40 & 273000  & 4.47  &  0.81  &  1.25 & 12.61 & -6.31 \\  
10:40--10:45 & 312000  & 4.25  &  0.8   &  1.75 & 14.32 & -8.02 \\  
10:45--10:50 & 340100  & 4.49  &  0.68  &  2. 5 & 15.46 & -14.32 \\  
10:50--10:55 & 328200  & 4.87  &  0.41  &  4. 5 & 20.62 & -14.89 \\  
10:55--11:00 & 290000  & 5.51  &  0.05  &  11. 5 & 10.88 & -12.65 \\  
11:25--11:30 & 272000  & 5.52  &  0.05  &  11. 8 & 14.32 & -16.04 \\  
11:55--12:00 & 230000  & 5.74  &  0.02  &  12. 1 & 6.3   & -21.19 \\  
12:25--12:30 & 195200  & 5.83  &  0.03  &  12. 5 & 8.59  & -23.49 \\  
12:55--13:00 & 193000  & 6.01  &  0.0   &  12. 6 & 6.28  & -25.78 \\  
13:25--13:30 & 187100  & 6.15  &  0.0   &  13. 1 & 6.87  & -29.22 \\  
13:55--14:00 & 182500  & 6.22  &  0.0   &  13. 5 & 6.91  & -32.65 \\  
14:55--15:00 & 178500  & 6.5   &  0.0   &  14. 2 & 8.59  & -35.52 \\  
15:55--16:00 & 165000  & 7.1   &  0.0   &  15.0  & 5.72  & -38.39 \\  

\hline                  
\end{tabular}
\end{table*}

\begin{figure}[H]
   \centering
   \includegraphics[width=1.2\textwidth]{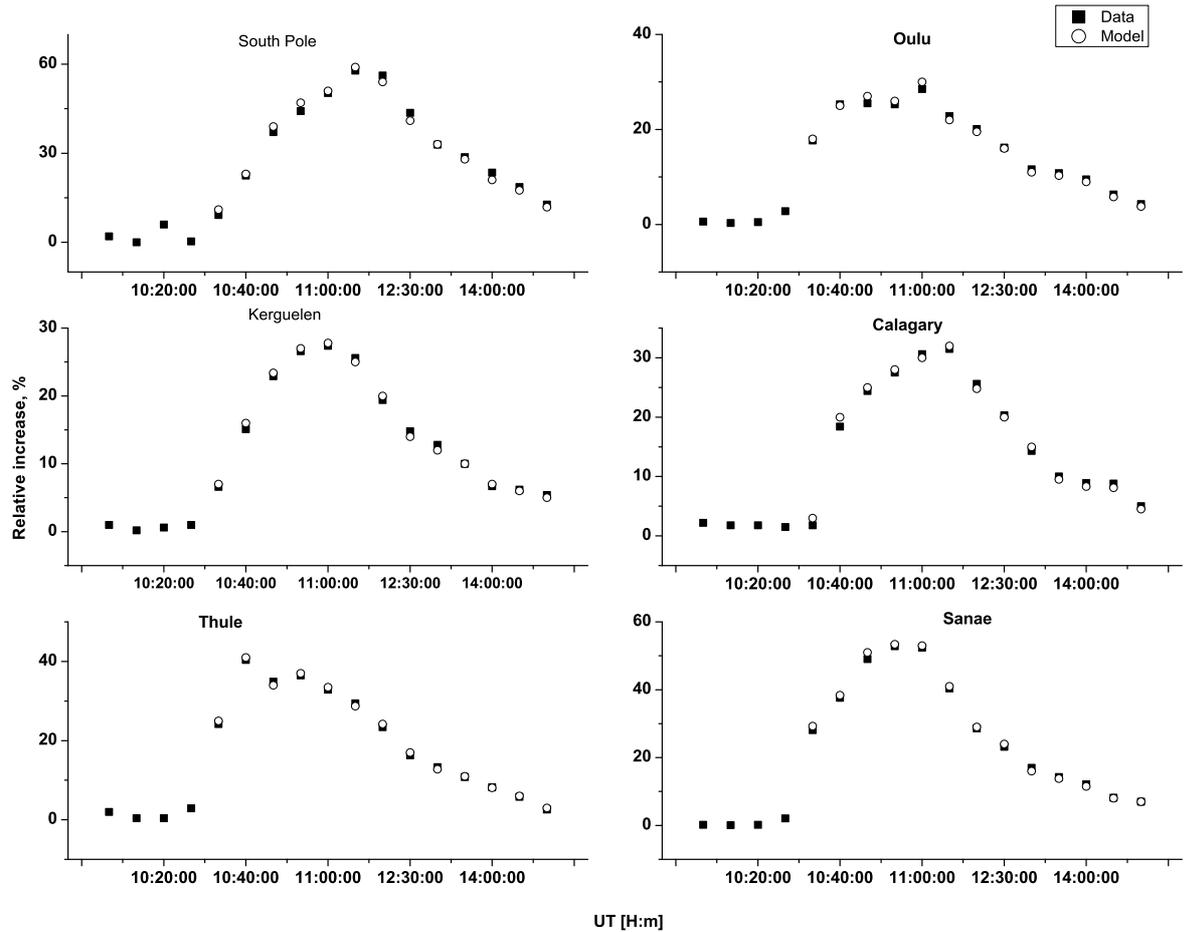}
      \caption{Modelled and observed responses of several NM stations during the GLE 59 on 14 July 2000. The quality of the fit for other stations is of the same order.
              }
         \label{Fig4}
   \end{figure}

\subsection{The GLE 70 on 13 December 2006}
The second event considered for analysis in this study is one of the strongest events of Solar Cycle 23. The month of December 2006 was at the declining phase of the solar cycle close to the minimum. However, on 13 December 2006, NOAA active region 10930, located at S06W26, triggered a X3.4/4 B solar flare which reached maximum at 2:40 UT. It was associated with Type II and Type IV radio bursts and a fast full-halo CME accompanied by a strong solar proton event \citep[for details see \textit{e.g.}][and references therein]{Gopalswamy2012, Moraal12}. The worldwide network of NMs recorded the event with maximum seen at Oulu NM $\sim$ 90 $\%$ at 5 min data) (Figure 5). It was classified as GLE 70. 

\begin{figure}[H]
   \centering
   \includegraphics[width=\textwidth]{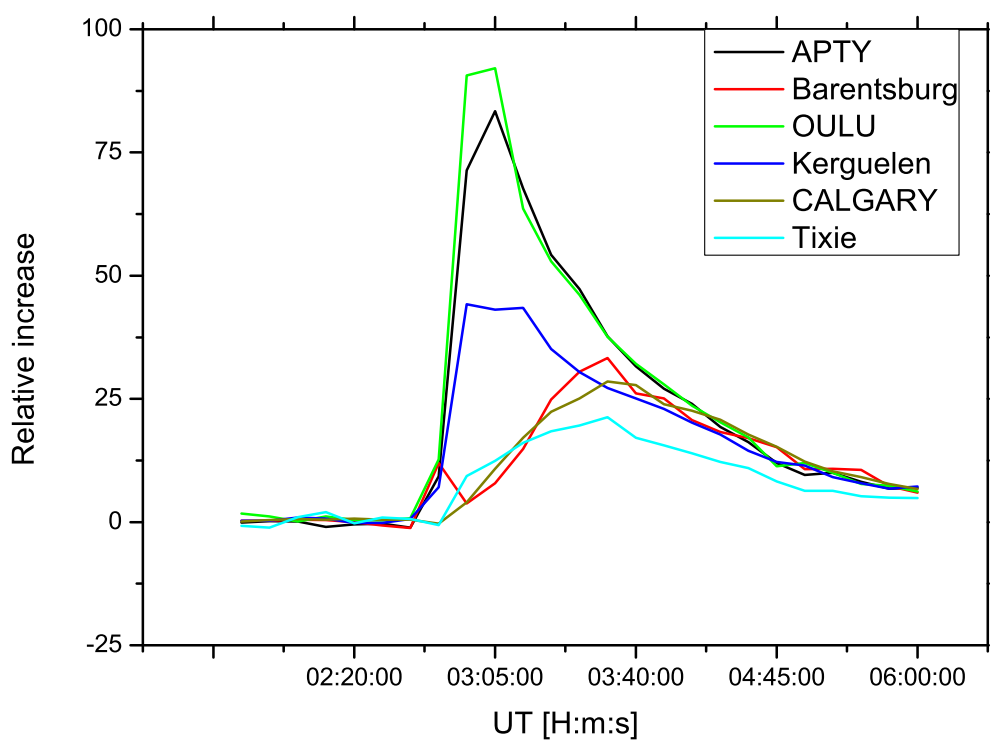}
      \caption{Time variation of 10-min data in Apatity, Barenstburg, Calgary, Kerguelen, Oulu and Tixie NMs relative increase during GLE 70 on 13 December 2006. 
              }
         \label{Fig5}
   \end{figure}

Similarly to some other events it was characterized by a large anisotropy in the initial phase \citep{But09, Vas08, Plainaki2009}. Here for the analysis we consider 5-min NM data retrieved from GLE database \citep{Usoskin2015a}. An illustration of several computed asymptotic cones of NMs used for the analysis is shown in Figure 6. The full list of NMs used in this analysis is given in Table 3. Similarly to the previous case, the computations are carried out with the MAGNETOCOSMICS code using Tsyganenko 1989 (external field model) and IGRF (internal field model) geomagnetic models, adjusted to the measured $K_{p}$ index and 2000 epoch. The derived rigidity spectra with the corresponding anisotropy characteristics are shown in Figure 7. During the event's onset, SEPs had a hard spectrum, strong anisotropy of a beam like SEP flux, which have been observed by NM stations with small pitch-angles. As an example, Oulu and Barentsburg NMs showed very different responses, despite their close geographic location and rigidity cut-offs. A summary of the derived spectral and angular characteristics for GLE 70 on 13 December 2006, as well as NM integration period and apparent source position is given in Table 2. 

\begin{figure}[H]
   \centering
   \includegraphics[width=\textwidth]{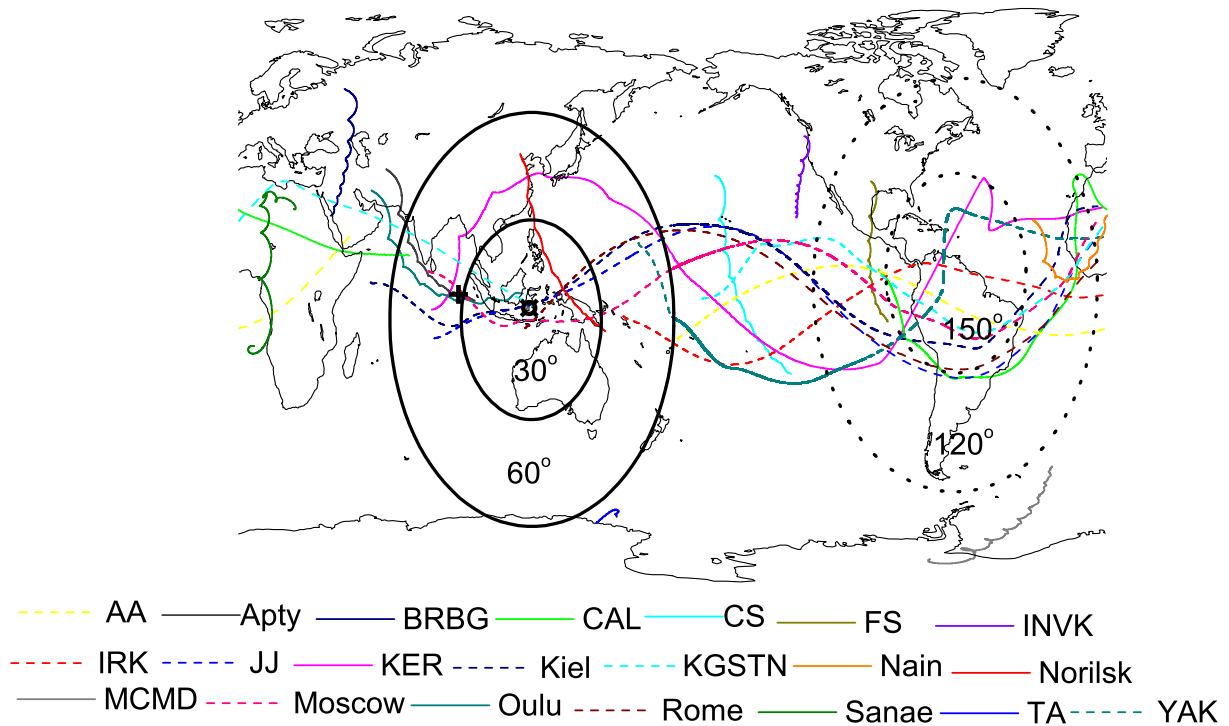}
      \caption{Calculated NM asymptotic directions during GLE 70 on 13 December 2006 at 03:00 UT. The cross represents the direction of interplanetary magnetic field (IMF) derived from the ACE satellite measurements at 03:00 UT. The small oval represents the derived apparent source position. The lines of equal pitch angles relative to the derived anisotropy axis are plotted for 30$^{\circ}$, 60$^{\circ}$, 120$^{\circ}$ and 150$^{\circ}$. The asymptotic directions of polar NMs are plotted with solid lines, while mid-latitude NMs are plotted with dashed lines.
              }
         \label{Fig6}
   \end{figure}

\begin{figure}[H]
   \centering
   \includegraphics[width=\textwidth]{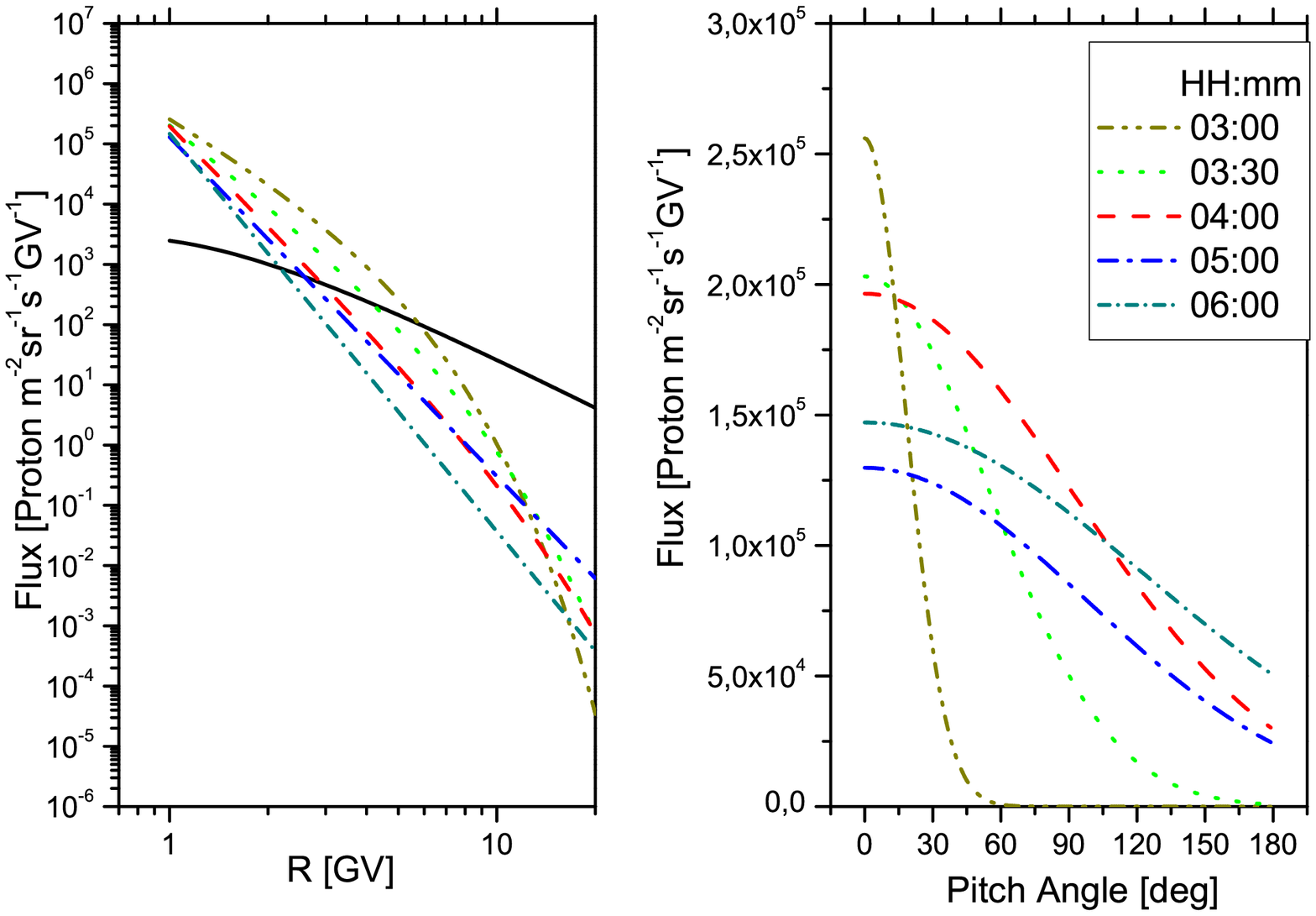}
      \caption{Derived rigidity spectra and pitch angle distributions of SEPs during the GLE 70. The SEP flux is according to Equation (2) \textit{i.e.} flux arriving from the Sun along the axis of symmetry. Time (UT) refers to the end of the corresponding five minute interval. The solid line in the left panel denote GCR flux.
              }
         \label{Fig3}
   \end{figure}

\begin{table*}[htp]
\caption{Derived spectral and angular characteristics for GLE 70 on 13 December 2006}             
\label{table:A2}      
\centering          
\begin{tabular}{c c c c c c c }     
    

\hline       
Integration interval [UT] & $J_{0}$ [m$^{-2}$ s$^{-1}$ sr$^{-1}$ GV$^{-1}$] & $\gamma$ & $\delta\gamma$ & $\sigma^{2}$ [rad $^{2}$]  & $\Psi$ [degrees]& $\Lambda$ [degrees]\\ 
\hline                    
03:00--03:05 & 256000  & 3.41  &  0.22  &  0.19 & -17 & 148 \\  
03:05--03:10 & 356500  & 3.71  &  0.3   &  0.28 & -16 & 154 \\  
03:10--03:15 & 216800  & 4.25  &  0.2   &  0.3  & -14 & 145 \\  
03:15--03:20 & 206000  & 4.31  &  0.1   &  0.85 & -13 & 160 \\  
03:20--03:25 & 205000  & 4.32  &  0.2   &  1.2  & -13 & 137 \\  
03:25--03:30 & 203100  & 4.43  &  0.11  &  1.77 & -11 & 131 \\  
03:30--03:35 & 200000  & 4.5   &  0.1   &  1.81 & -10 & 126 \\  
03:35--03:40 & 200000  & 4.75  &  0.08  &  1.85 & -10 & 128 \\  
03:40--03:45 & 204900  & 5.2   &  0.1   &  2.01 & -7  & 122 \\  
03:45--03:50 & 202500  & 5.46  &  0.12  &  2.3  & -5  & 120 \\  
03:50--03:55 & 201800  & 5.48  &  0.1   &  3.5  & -5  & 117 \\  
03:55--04:00 & 196500  & 5.52  &  0.05  &  5.19 & -1  & 120 \\
04:10--04:15 & 185203  & 5.62  &  0.04  &  5.8  & 0.1 & 118 \\ 
04:25--04:30 & 150000  & 5.69  &  0.001 &  5.68 & 5   & 115 \\ 
04:40--04:45 & 135920  & 5.65  &  0     &  6.47 & 10  & 111 \\ 
04:55--05:00 & 129800  & 5.63  &  0     &  5.87 & 13  & 110 \\
05:10--05:15 & 145325  & 6.15  &  0     &  6.63 & 15  & 108 \\ 
05:25--05:30 & 135320  & 6.12  &  0     &  6.68 & 14  & 107 \\  
05:40--05:45 & 125050  & 6.19  &  0     &  7.1  & 14  & 105 \\ 
05:55--06:00 & 147150  & 6.6   &  0     &  9.2  & 15  & 95  \\
\hline                  
\end{tabular}
\end{table*}

The quality of the modelling is demonstrated in Figure 8, where the model and observed responses of several NMs are compared. The quality of the fit is similar for the other NM stations. Similarly to Figure 4, the X axis (Time) in Figure 8 is not uniform.

The analysis similar to the previous one shows an achieved maximal relative difference of about 3-6 $\%$ between modelled and observed NM relative increase. The full list of NMs used for the analysis of both events is given in Table 3.

\begin{figure}[H]
   \centering
   \includegraphics[width=1.2\textwidth]{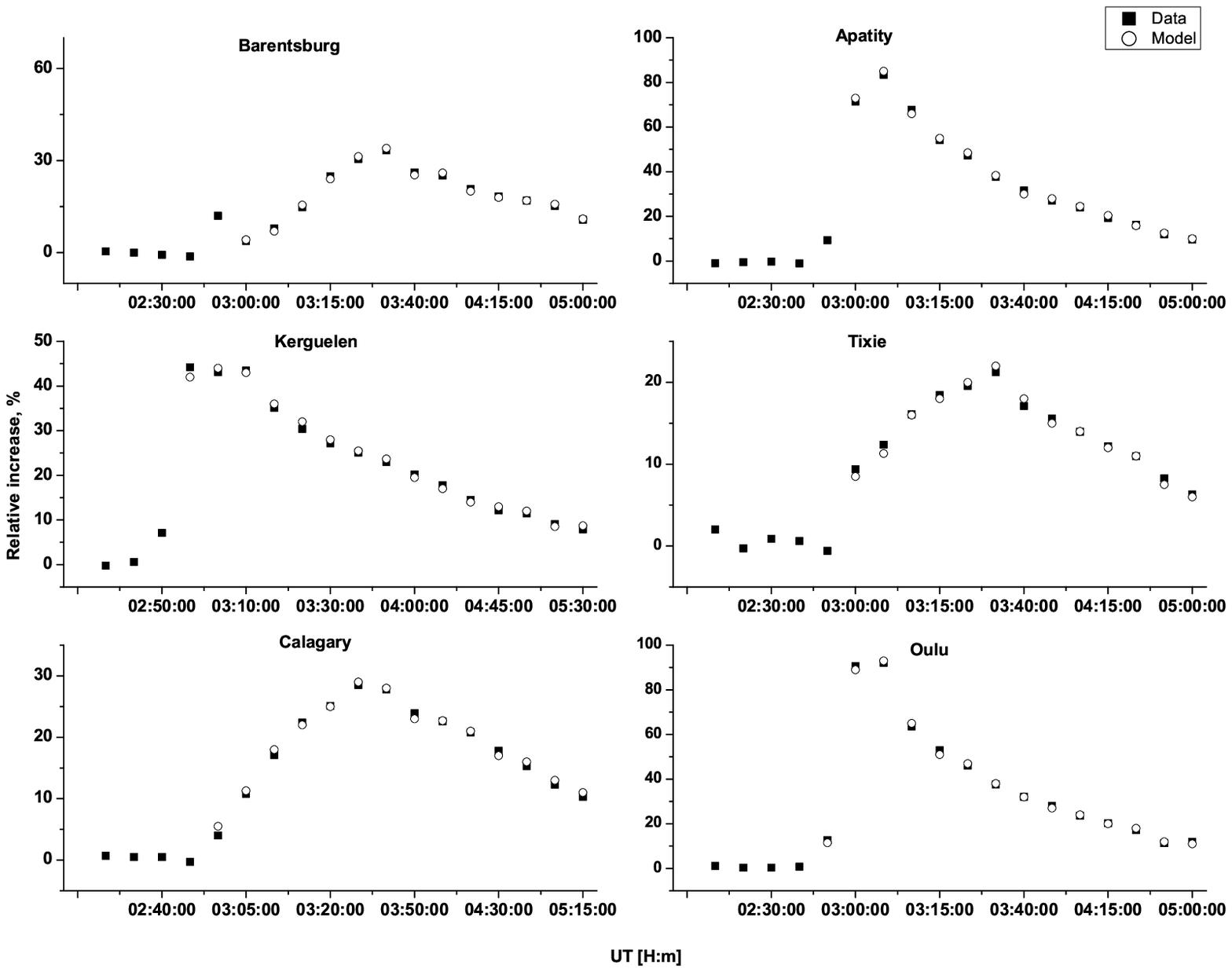}
      \caption{Modelled and observed responses of six NM stations during the GLE70 on 13 December 2006. The quality of the modelled responses for other stations is of the same order.
              }
         \label{Fig4}
   \end{figure}

\begin{table*}[htp]
\caption{Neutron monitors with corresponding geomagnetic rigidity cut-offs used in the analysis. Cross (null) in the last columns denote NMs used (not used) for the analysis of the event.}             
\label{table:b2}      
\centering          
\begin{tabular}{c c c c c c c }     
\hline\hline       

Station & latitude [deg] & Longitude [deg] &  $P_{c}$ [GV] & Altitude [m] & GLE 59 & GLE 70 \\ 
\hline                    
   Alma Ata (AA) & 43.25  & 76.92  & 6.67 & 3340  & x & x \\  
   Apatity (Apty)& 67.55  & 33.33  & 0.48 &  177  & x & x \\
   Barentsburg (BRBG) & 78.03  & 14.13  & 0 & 70  & 0 & x \\
   Calgary (Cal) & 51.08 & 245.87  & 1.04  &  1128 & x & x \\
   Cape Schmidt (CS) & 68.92 & 180.53  & 0.41 &  0 & 0 & x \\
   Forth Smith (FS)  & 60.02 & 248.07  & 0.25 &  0 & 0 & x \\
   Goose Bay (GB) & 23.27 & 299.60  & 0.52  &  46  & x & 0  \\
   Hermanus (HRMS)  & -34.42 & 19.22 & 4.90  & 26  & x & x  \\
   Hobart (HBRT)  & -42.92 & 147.24 & 1.88  &  0   & x & 0  \\
   Inuvik (INVK)     & 68.35  & 226.28  & 0.16  &  21  & x & x  \\
   Irkutsk (IRK)  & 52.58 & 104.02 & 3.23 &  435 & x & x \\
   Jungfraujoch (JJ) & 46.55 & 7.98 & 4.46 & 3476 & x & x \\
   Kerguelen (Ker) & -49.35 & 70.25 & 1.01 & 33 & x & x \\
   Kiel & 54.33 & 10.13 & 2.22 & 54 & x & x \\
   Kingston (KGSTN) & -42.99 & 147.29 & 1.75 & 65 & x & x\\
   Lomnicky \v{S}tit (LS) & 49.2 & 20.22 & 3.72 & 2634 & x & x\\ 
   Magadan (MAG) & 60.12 & 151.02 & 1.84 & 220 & x & x\\
   Mawson (MWSN) & -67.60 & 62.88 & 0.22 & 0 & x & x \\
   McMurdo (MCMD) & -77.85 & 166.72 & 0 & 48 & x & x\\
   Moscow (MOS) & 55.47 & 37.32 & 2.13 & 200 & x & x\\
   Nain & 56.55 & 298.32 & 0.28 & 0 & 0 & x \\
   Newark (NWRK)& 39.70 & 284.30 & 1.97 & 50 & x & 0 \\
   Norilsk & 69.26 & 88.05 & 0.52 & 0 & 0 & x \\
   Oulu & 65.05 & 25.47 & 0.69 & 15 & x & x\\
   Peawanuck (PWNC) & 54.98 & 274.56 & 0.16 & 52 & 0 & x\\
   Rome & 41.86 & 12.47 & 6.19 & 60 & 0 & x \\
   Sanae & -71.67 & 357.15 & 0.56 & 856 & x & 0 \\
   South Pole (SP) & -90.00 & 0.0 & 0 & 2820 & x & 0 \\
   Terre Adelie (TA) & -66.67 & 140.02 & 0 & 45 & x & x\\
   Thule & 76.60 & 291.2 & 0.1 & 260 & x & 0\\
   Tixie (TB) & 71.60 & 128.90 & 0.53 & 0 & x & x\\
   Yakutsk (YAK) & 62.03 & 129.73 & 1.64 & 105 & x & x \\
   
\hline                  
\end{tabular}
\end{table*}

\section{Discussion and Conclusions}
The two major events GLE 59 and GLE 70 considered for study in this work occurred at different solar activity conditions. They were also quite different in increase profiles. The GLE 70 depicted a very sharp impulsive-type increase, while GLE 59 had a wider time profile typical for gradual events. This feature may be explained by the position of the flare at the solar disk \citep{Duldig93, Cramp97}.

The events were characterized by relatively strong anisotropy during the initial phase, which decreased rapidly over the following 30 minutes for GLE 59, accordingly 50-60 minutes for GLE 70. The strong anisotropy of SEPs during both event onsets indicates focused transport conditions in the interplanetary medium. In addition, we studied the likelihood of bidirectional arrival of SEPs, namely by modelling the NM response assuming a two flux event and a complicated shape of pitch angle distribution (see \citet{Mishev14c}) similarly to \citet{Vas06b} who found a clear signature of bidirectional flux for the GLE 69 on 20 January 2005. In this case analysis we found no evidence of particles arriving from other but sunward directions. The rapid increase of NM response of stations in the antisunward direction such as Tixie for GLE 59 is most-likely due to scattering processes \citep{Bieber02}.

The particle fluences (rigidity, time and angle integrated particle flux) of both events are compared (Figure 9a). The particle fluence during the Bastille day event is greater than during the GLE 70 event. This is consistent with recent findings based on both satellite-borne and NMs data analysis \citep{Tylka09}. The fluence during Bastille day event is compared with recent estimations (Figure 9b) \citep{Tylka09}. The observed discrepancy is most likely due to different reconstruction methods and model assumptions. In \citep{Tylka09} a simplified analysis of NM data is used and different spectral shape, namely band function is employed. 

A reasonable agreement is achieved for GLE 70 event (Figure 9c). In addition, on Figure 9c is compared the particle fluence using the reconstructions by \citet{Vas08}, where a discrepancy in particle fluence is observed, but not in derived rigidity spectra. 

\begin{figure}[H]
   \centering
   \includegraphics[width=\textwidth]{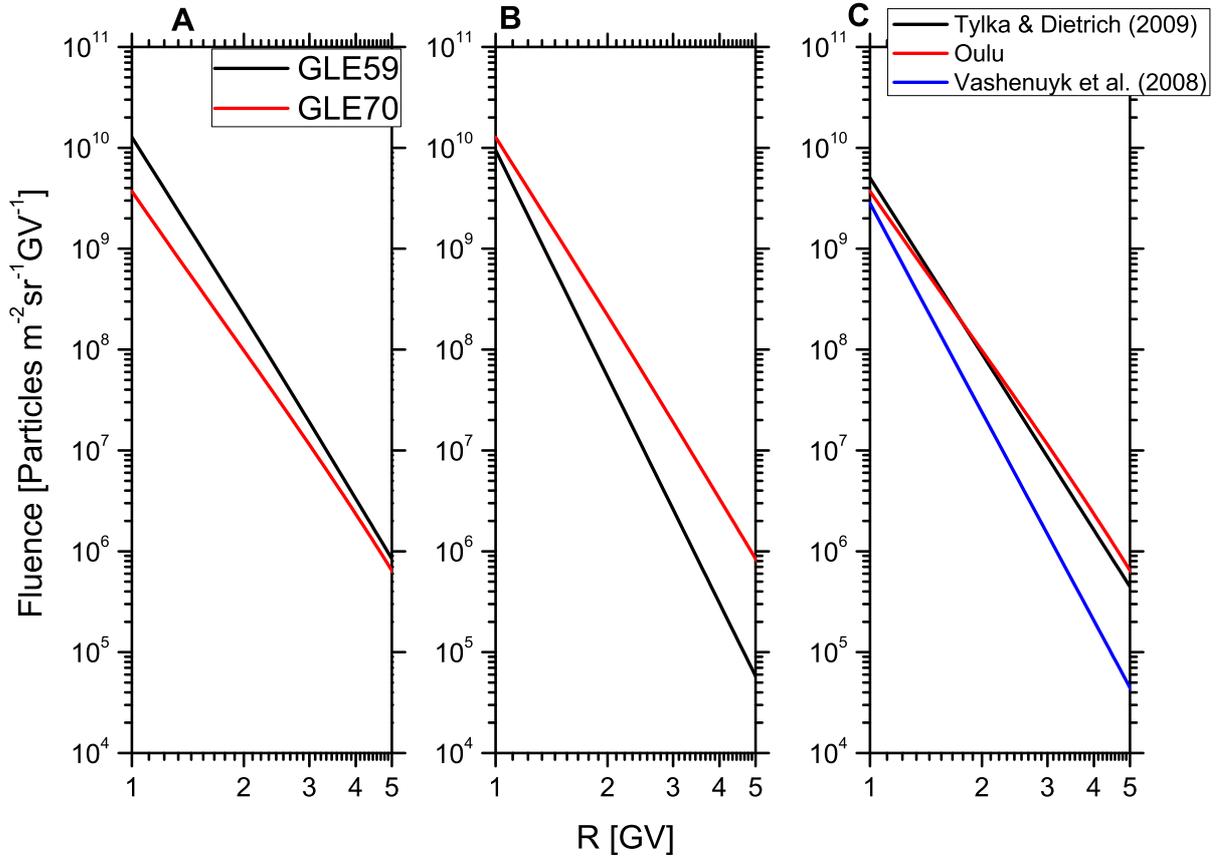}
      \caption{ a) Computed fluence of SEPs during GLE 59 on 14 July 2000 and GLE 70  on 13 December 2006 as denoted in the legend in panel A. The SEPs fluence is time- and angle- integrated throughout the event(s); b) Computed fluence of SEPs during GLE 59 on 14 July 2000 compared with previous estimations \citep{Tylka09} as denoted in the legend in panel C; c) Computed fluence of SEPs during GLE 70 on 13 December 2006 compared with previous estimations \citep{Vas08, Tylka09} as denoted in the legend in panel C;  
              }
         \label{Fig9}
   \end{figure}

In the work presented here, we performed a detailed modelling of spectral and angular characteristics of high energy SEPs in the vicinity of Earth during the 
the Bastille Day 2000 (GLE 59) event and GLE 70 on 13 December 2006. 

The CR particles are the main source of ionization in the troposphere and stratosphere resulting form the induced nuclear-electromagnetic-muon cascade and secondary particle energy loss \citep{ Usoskin2006, Bazilevskaya08, Velinov2013, Mironova2015}. The derived characteristics during the GLEs are useful for further space weather/space climate applications, namely the estimation of ion production rate and ionization effect in the atmosphere similarly to \citep[see \textit{e.g.}][]{Bazilevskaya08, Uso11b, Zigman2014, Mishev2015b, Mishev2015c, Mishev2015d}.



\begin{thebibliography}{68}
\ifx\bisbn     \undefined \def\bisbn  #1{ISBN #1}\fi
\ifx\binits    \undefined \def\binits#1{#1}\fi
\ifx\bauthor   \undefined \def\bauthor#1{#1}\fi
\ifx\batitle   \undefined \def\batitle#1{#1}\fi
\ifx\bjtitle   \undefined \def\bjtitle#1{\textit{#1}}\fi
\ifx\bvolume   \undefined \def\bvolume#1{\textbf{#1}}\fi
\ifx\byear     \undefined \def\byear#1{#1}\fi
\ifx\bissue    \undefined \def\bissue#1{#1}\fi
\ifx\bfpage    \undefined \def\bfpage#1{#1}\fi
\ifx\blpage    \undefined \def\blpage #1{#1}\fi
\ifx\burl      \undefined \def\burl#1{\textsf{#1}}\fi
\ifx\href      \undefined \def\href#1#2{\textsf{#2}}\fi
\ifx\betal     \undefined \def\betal{\textit{et al.}}\fi
\ifx\bctitle   \undefined \def\bctitle#1{#1}\fi
\ifx\beditor   \undefined \def\beditor#1{#1}\fi
\ifx\bbtitle   \undefined \def\bbtitle#1{\textit{#1}}\fi
\ifx\bedition  \undefined \def\bedition#1{#1}\fi
\ifx\bseriesno \undefined \def\bseriesno#1{\textbf{#1}}\fi
\ifx\blocation \undefined \def\blocation#1{#1}\fi
\ifx\bsertitle \undefined \def\bsertitle#1{\textit{#1}}\fi
\ifx\bsnm      \undefined \def\bsnm#1{#1}\fi
\ifx\bsuffix   \undefined \def\bsuffix#1{#1}\fi
\ifx\bparticle \undefined \def\bparticle#1{#1}\fi
\ifx\barticle  \undefined \def\barticle#1{}\fi
\ifx\binstitute  \undefined \def\binstitute#1{#1}\fi
\ifx\bpublisher  \undefined \def\bpublisher#1{#1}\fi
\ifx\doiurl    \undefined
  \def\doiurl#1{\href{http://dx.doi.org/#1}{\textsf{DOI}}}\fi
\ifx\arxivurl  \undefined
  \def\arxivurl#1{\href{http://arxiv.org/abs/#1}{\textsf{arXiv}}}\fi
\ifx\adsurl    \undefined
  \def\adsurl#1{\href{http://adsabs.harvard.edu/abs/#1}{\textsf{ADS}}}\fi
\ifx\botherref \undefined \def\botherref#1{}\fi
\ifx\url       \undefined \def\url#1{\textsf{#1}}\fi
\ifx\bchapter  \undefined \def\bchapter#1{}\fi
\ifx\bbook     \undefined \def\bbook#1{}\fi
\ifx\bcomment  \undefined \def\bcomment#1{#1}\fi
\ifx\oauthor   \undefined \def\oauthor#1{#1}\fi
\ifx\citeauthoryear \undefined\def \citeauthoryear#1{#1}\fi
\def\endbibitem {}
\ifx\bconflocation  \undefined \def\bconflocation#1{#1} \fi

\bibitem[\protect\citeauthoryear{Andriopoulou
  \textit{et~al.}}{2011a}]{Andriopoulou2011a}
\begin{barticle}
\bauthor{\bsnm{Andriopoulou}, \binits{M.}},
\bauthor{\bsnm{Mavromichalaki}, \binits{H.}},
\bauthor{\bsnm{Plainaki}, \binits{C.}},
\bauthor{\bsnm{Belov}, \binits{A.}},
\bauthor{\bsnm{Eroshenko}, \binits{E.}}:
\byear{2011}a,
\batitle{Intense ground-level enhancements of solar cosmic rays during the last
  solar cycles}.
\bjtitle{Solar Physics}
\bvolume{269}(\bissue{1}),
\bfpage{155}.

\end{barticle}
\endbibitem

\bibitem[\protect\citeauthoryear{Andriopoulou
  \textit{et~al.}}{2011b}]{Andriopoulou2011b}
\begin{barticle}
\bauthor{\bsnm{Andriopoulou}, \binits{M.}},
\bauthor{\bsnm{Mavromichalaki}, \binits{H.}},
\bauthor{\bsnm{Preka-Papadema}, \binits{P.}},
\bauthor{\bsnm{Plainaki}, \binits{C.}},
\bauthor{\bsnm{Belov}, \binits{A.}},
\bauthor{\bsnm{Eroshenko}, \binits{E.}}:
\byear{2011}b,
\batitle{Solar activity and the associated ground level enhancements of solar
  cosmic rays during solar cycle 23}.
\bjtitle{Astrophysics and Space Sciences Transactions}
\bvolume{7}(\bissue{4}),
\bfpage{439}.

\end{barticle}
\endbibitem

\bibitem[\protect\citeauthoryear{Aschwanden}{2012}]{Aschwanden12}
\begin{barticle}
\bauthor{\bsnm{Aschwanden}, \binits{M.}}:
\byear{2012},
\batitle{GeV particle acceleration in solar flares and ground level enhancement
  (GLE) events}.
\bjtitle{Space Science Reviews}
\bvolume{171}(\bissue{1-4}),
\bfpage{3}.

\end{barticle}
\endbibitem

\bibitem[\protect\citeauthoryear{Aster, Borchers, and
  Thurber}{2005}]{Aster2005}
\begin{bbook}
\bauthor{\bsnm{Aster}, \binits{R.C.}},
\bauthor{\bsnm{Borchers}, \binits{B.}},
\bauthor{\bsnm{Thurber}, \binits{C.H.}}:
\byear{2005},
\bbtitle{Parameter estimation and inverse problems},
\bpublisher{Elsevier},
\blocation{New York}.
\bisbn{0-12-065604-3}.
\end{bbook}
\endbibitem

\bibitem[\protect\citeauthoryear{Bazilevskaya
  \textit{et~al.}}{2008}]{Bazilevskaya08}
\begin{barticle}
\bauthor{\bsnm{Bazilevskaya}, \binits{G.A.}},
\bauthor{\bsnm{Usoskin}, \binits{I.G.}},
\bauthor{\bsnm{Fl{\"u}ckiger}, \binits{E.O.}},
\bauthor{\bsnm{Harrison}, \binits{R.G.}},
\bauthor{\bsnm{Desorgher}, \binits{L.}},
\bauthor{\bsnm{B{\"u}tikofer}, \binits{B.}},
\bauthor{\bsnm{Krainev}, \binits{M.B.}},
\bauthor{\bsnm{Makhmutov}, \binits{V.S.}},
\bauthor{\bsnm{Stozhkov}, \binits{Y.I.}},
\bauthor{\bsnm{Svirzhevskaya}, \binits{A.K.}},
\bauthor{\bsnm{Svirzhevsky}, \binits{N.S.}},
\bauthor{\bsnm{Kovaltsov}, \binits{G.A.}}:
\byear{2008},
\batitle{Cosmic ray induced ion production in the atmosphere}.
\bjtitle{Space Science Reviews}
\bvolume{137},
\bfpage{149}.

\end{barticle}
\endbibitem

\bibitem[\protect\citeauthoryear{Bieber and Evenson}{1995}]{Bieber95}
\begin{bchapter}
\bauthor{\bsnm{Bieber}, \binits{J.W.}},
\bauthor{\bsnm{Evenson}, \binits{P.A.}}:
\byear{1995},
\bctitle{Spaceship Earth - an optimized network of neutron monitors}.
In: \bbtitle{Proc. of 24th ICRC Rome, Italy, 28 Auygust - 8 September 1995}
\bseriesno{4},
\bfpage{1316}.
\end{bchapter}
\endbibitem

\bibitem[\protect\citeauthoryear{Bieber \textit{et~al.}}{2002}]{Bieber02}
\begin{barticle}
\bauthor{\bsnm{Bieber}, \binits{J.W.}},
\bauthor{\bsnm{Droge}, \binits{W.}},
\bauthor{\bsnm{Evenson}, \binits{P.A.}},
\bauthor{\bsnm{Pyle}, \binits{K.R.}},
\bauthor{\bsnm{Ruffolo}, \binits{D.}},
\bauthor{\bsnm{Pinsook}, \binits{U.}},
\bauthor{\bsnm{Tooprakai}, \binits{P.}},
\bauthor{\bsnm{Rujiwarodom}, \binits{M.}},
\bauthor{\bsnm{Khumlumlert}, \binits{T.}},
\bauthor{\bsnm{Krucker}, \binits{S.}}:
\byear{2002},
\batitle{Energetic particle observations during the 2000 July 14 solar event}.
\bjtitle{Astrophysical Journal}
\bvolume{567}(\bissue{1}),
\bfpage{622}.

\end{barticle}
\endbibitem

\bibitem[\protect\citeauthoryear{Bieber \textit{et~al.}}{2013}]{Bieber2013}
\begin{barticle}
\bauthor{\bsnm{Bieber}, \binits{J.W.}},
\bauthor{\bsnm{Clem}, \binits{J.}},
\bauthor{\bsnm{Evenson}, \binits{P.}},
\bauthor{\bsnm{Pyle}, \binits{R.}},
\bauthor{\bsnm{Sáiz}, \binits{A.}},
\bauthor{\bsnm{Ruffolo}, \binits{D.}}:
\byear{2013},
\batitle{Giant ground level enhancement of relativistic solar protons on 2005
  January 20. i. spaceship Earth observations}.
\bjtitle{Astrophysical Journal}
\bvolume{771}(\bissue{2}).

\end{barticle}
\endbibitem

\bibitem[\protect\citeauthoryear{Bombardieri \textit{et~al.}}{2006}]{Bom06}
\begin{barticle}
\bauthor{\bsnm{Bombardieri}, \binits{D.J.}},
\bauthor{\bsnm{Duldig}, \binits{M.L.}},
\bauthor{\bsnm{Michael}, \binits{K.J.}},
\bauthor{\bsnm{Humble}, \binits{J.E.}}:
\byear{2006},
\batitle{Relativistic proton production during the 2000 July 14 solar event:
  The case for multiple source mechanisms}.
\bjtitle{Astrophysical Journal}
\bvolume{644}(\bissue{1}),
\bfpage{565}.

\end{barticle}
\endbibitem

\bibitem[\protect\citeauthoryear{Bombardieri
  \textit{et~al.}}{2008}]{Bombardier2008}
\begin{barticle}
\bauthor{\bsnm{Bombardieri}, \binits{D.J.}},
\bauthor{\bsnm{Duldig}, \binits{M.L.}},
\bauthor{\bsnm{Humble}, \binits{J.E.}},
\bauthor{\bsnm{Michael}, \binits{K.J.}}:
\byear{2008},
\batitle{An improved model for relativistic solar proton acceleration applied
  to the 2005 January 20 and earlier events}.
\bjtitle{Astrophysical Journal}
\bvolume{682}(\bissue{2}),
\bfpage{1315}.

\end{barticle}
\endbibitem

\bibitem[\protect\citeauthoryear{B{\"u}tikofer and
  Fl{\"u}ckiger}{2013}]{But13a}
\begin{barticle}
\bauthor{\bsnm{B{\"u}tikofer}, \binits{R.}},
\bauthor{\bsnm{Fl{\"u}ckiger}, \binits{E.O.}}:
\byear{2013},
\batitle{Differences in published characteristics of GLE 60 and their
  consequences on computed radiation dose rates along selected flight paths}.
\bjtitle{Journal of Physics: Conference Series}
\bvolume{409}(\bissue{1}),
\bfpage{012166}.

\end{barticle}
\endbibitem

\bibitem[\protect\citeauthoryear{B{\"u}tikofer \textit{et~al.}}{2009}]{But09}
\begin{barticle}
\bauthor{\bsnm{B{\"u}tikofer}, \binits{R.}},
\bauthor{\bsnm{Fl{\"u}ckiger}, \binits{E.O.}},
\bauthor{\bsnm{Desorgher}, \binits{L.}},
\bauthor{\bsnm{Moser}, \binits{M.R.}},
\bauthor{\bsnm{Pirard}, \binits{B.}}:
\byear{2009},
\batitle{The solar cosmic ray ground-level enhancements on 20 January 2005 and
  13 December 2006}.
\bjtitle{Advances in Space Research}
\bvolume{43}(\bissue{4}),
\bfpage{499}.

\end{barticle}
\endbibitem

\bibitem[\protect\citeauthoryear{B{\"u}tikofer \textit{et~al.}}{2013}]{But13b}
\begin{bchapter}
\bauthor{\bsnm{B{\"u}tikofer}, \binits{R.}},
\bauthor{\bsnm{Fl{\"u}ckiger}, \binits{E.O.}},
\bauthor{\bsnm{Balabin}, \binits{Y.}},
\bauthor{\bsnm{Belov}, \binits{A.}}:
\byear{2013},
\bctitle{The reliability of GLE analysis based on neutron monitor data - a
  critical review}.
In: \bbtitle{Proc. of 33th ICRC Rio de Janeiro, Brazil, 2 -9 July 2013},
\bfpage{0863}.
\end{bchapter}
\endbibitem

\bibitem[\protect\citeauthoryear{Caballero-Lopez and Moraal}{2004}]{Cab04}
\begin{barticle}
\bauthor{\bsnm{Caballero-Lopez}, \binits{R.A.}},
\bauthor{\bsnm{Moraal}, \binits{H.}}:
\byear{2004},
\batitle{Limitations of the force field equation to describe cosmic ray
  modulation}.
\bjtitle{Journal of Geophysical Research}
\bvolume{109},
\bfpage{A01101}.

\end{barticle}
\endbibitem

\bibitem[\protect\citeauthoryear{Clem and Dorman}{2000}]{Cle00}
\begin{barticle}
\bauthor{\bsnm{Clem}, \binits{J.}},
\bauthor{\bsnm{Dorman}, \binits{L.}}:
\byear{2000},
\batitle{Neutron monitor response functions}.
\bjtitle{Space Science Reviews}
\bvolume{93},
\bfpage{335}.
\end{barticle}
\endbibitem

\bibitem[\protect\citeauthoryear{Cliver, Kahler, and Reames}{2004}]{Cliver04}
\begin{barticle}
\bauthor{\bsnm{Cliver}, \binits{E.W.}},
\bauthor{\bsnm{Kahler}, \binits{S.W.}},
\bauthor{\bsnm{Reames}, \binits{D.V.}}:
\byear{2004},
\batitle{Coronal shocks and solar energetic proton events}.
\bjtitle{Astrophysical Journal}
\bvolume{605},
\bfpage{902}.
\end{barticle}
\endbibitem

\bibitem[\protect\citeauthoryear{Cooke \textit{et~al.}}{1991}]{Cook91}
\begin{barticle}
\bauthor{\bsnm{Cooke}, \binits{D.J.}},
\bauthor{\bsnm{Humble}, \binits{J.E.}},
\bauthor{\bsnm{Shea}, \binits{M.A.}},
\bauthor{\bsnm{Smart}, \binits{D.F.}},
\bauthor{\bsnm{Lund}, \binits{N.}},
\bauthor{\bsnm{Rasmussen}, \binits{I.L.}},
\bauthor{\bsnm{Byrnak}, \binits{B.}},
\bauthor{\bsnm{Goret}, \binits{P.}},
\bauthor{\bsnm{Petrou}, \binits{N.}}:
\byear{1991},
\batitle{On cosmic-ray cutoff terminology}.
\bjtitle{Il Nuovo Cimento C}
\bvolume{14}(\bissue{3}),
\bfpage{213}.
\end{barticle}
\endbibitem

\bibitem[\protect\citeauthoryear{Cramp, Humble, and Duldig}{1995}]{Cramp95}
\begin{bchapter}
\bauthor{\bsnm{Cramp}, \binits{J.L.}},
\bauthor{\bsnm{Humble}, \binits{J.E.}},
\bauthor{\bsnm{Duldig}, \binits{M.L.}}:
\byear{1995},
\bctitle{The cosmic ray ground-level enhancement of 24 October 1989}.
In: \bbtitle{Proceedings Astronomical Society of Australia}
\bseriesno{11},
\bfpage{28}.
\end{bchapter}
\endbibitem

\bibitem[\protect\citeauthoryear{Cramp \textit{et~al.}}{1997}]{Cramp97}
\begin{barticle}
\bauthor{\bsnm{Cramp}, \binits{J.L.}},
\bauthor{\bsnm{Duldig}, \binits{M.L.}},
\bauthor{\bsnm{Fl{\"u}ckiger}, \binits{E.O.}},
\bauthor{\bsnm{Humble}, \binits{J.E.}},
\bauthor{\bsnm{Shea}, \binits{M.A.}},
\bauthor{\bsnm{Smart}, \binits{D.F.}}:
\byear{1997},
\batitle{The October 22, 1989,solar cosmic enhancement: ray an analysis the
  anisotropy spectral characteristics}.
\bjtitle{Journal of Geophysical Research}
\bvolume{102}(\bissue{A11}),
\bfpage{24 237}.
\end{barticle}
\endbibitem

\bibitem[\protect\citeauthoryear{Debrunner and Brunberg}{1968}]{Debrunner68}
\begin{barticle}
\bauthor{\bsnm{Debrunner}, \binits{H.}},
\bauthor{\bsnm{Brunberg}, \binits{E.}}:
\byear{1968},
\batitle{Monte Carlo calculation of nucleonic cascade in the atmosphere}.
\bjtitle{Canadian Journal of Physics}
\bvolume{46},
\bfpage{1069}.
\end{barticle}
\endbibitem

\bibitem[\protect\citeauthoryear{Debrunner \textit{et~al.}}{1988}]{Deb88}
\begin{barticle}
\bauthor{\bsnm{Debrunner}, \binits{H.}},
\bauthor{\bsnm{Fl{\"u}ckiger}, \binits{E.O.}},
\bauthor{\bsnm{Gradel}, \binits{H.}},
\bauthor{\bsnm{Lockwood}, \binits{J.A.}},
\bauthor{\bsnm{McGuire}, \binits{R.E.}}:
\byear{1988},
\batitle{Observations related to the acceleration, injection, and
  interplanetary propagation of energetic protons during the solar cosmic ray
  event on February 16, 1984}.
\bjtitle{Journal of Geophysical Research}
\bvolume{93}(\bissue{A7}),
\bfpage{7206}.
\end{barticle}
\endbibitem

\bibitem[\protect\citeauthoryear{Dennis and Schnabel}{1996}]{Den83}
\begin{bbook}
\bauthor{\bsnm{Dennis}, \binits{J.E.}},
\bauthor{\bsnm{Schnabel}, \binits{R.B.}}:
\byear{1996},
\bbtitle{Numerical methods for unconstrained optimization and nonlinear
  equations},
\bpublisher{Prentice-Hall},
\blocation{Englewood Cliffs}.
\bisbn{13-978-0-898713-64-0}.
\end{bbook}
\endbibitem

\bibitem[\protect\citeauthoryear{Desorgher \textit{et~al.}}{2005}]{Des05}
\begin{barticle}
\bauthor{\bsnm{Desorgher}, \binits{L.}},
\bauthor{\bsnm{Fl{\"u}ckiger}, \binits{E.O.}},
\bauthor{\bsnm{Gurtner}, \binits{M.}},
\bauthor{\bsnm{Moser}, \binits{M.R.}},
\bauthor{\bsnm{B{\"u}tikofer}, \binits{R.}}:
\byear{2005},
\batitle{A Geant 4 code for computing the interaction of cosmic rays with the
  earth's atmosphere}.
\bjtitle{Internationl Journal of Modern Physics A}
\bvolume{20}(\bissue{A11}),
\bfpage{6802}.

\end{barticle}
\endbibitem

\bibitem[\protect\citeauthoryear{Desorgher
  \textit{et~al.}}{2009}]{Desorgher2009}
\begin{barticle}
\bauthor{\bsnm{Desorgher}, \binits{L.}},
\bauthor{\bsnm{Kudela}, \binits{K.}},
\bauthor{\bsnm{Fl{\"u}ckiger}, \binits{E.O.}},
\bauthor{\bsnm{B{\"u}tikofer}, \binits{R.}},
\bauthor{\bsnm{Storini}, \binits{M.}},
\bauthor{\bsnm{Kalegaev}, \binits{V.}}:
\byear{2009},
\batitle{Comparison of earth's magnetospheric magnetic field models in the
  context of cosmic ray physics}.
\bjtitle{Acta Geophysica}
\bvolume{57}(\bissue{1}),
\bfpage{75}.

\end{barticle}
\endbibitem

\bibitem[\protect\citeauthoryear{Dorman}{2004}]{Dorman04}
\begin{bbook}
\bauthor{\bsnm{Dorman}, \binits{L.}}:
\byear{2004},
\bbtitle{Cosmic rays in the Earth's atmosphere and underground},
\bpublisher{Kluwer Academic Publishers},
\blocation{Dordrecht}.
\bisbn{1-4020-2071-6}.
\end{bbook}
\endbibitem

\bibitem[\protect\citeauthoryear{Dorman}{2006}]{Dor06}
\begin{bbook}
\bauthor{\bsnm{Dorman}, \binits{L.}}:
\byear{2006},
\bbtitle{Cosmic ray interactions, propagation, and acceleration in space
  plasmas},
\bpublisher{Astrophysics and Space Science Library 339, Springer},
\blocation{Dordrecht}.
\bisbn{13-978-1-4020-5100-5}.
\end{bbook}
\endbibitem

\bibitem[\protect\citeauthoryear{Dryer \textit{et~al.}}{2001}]{Dryer2001}
\begin{barticle}
\bauthor{\bsnm{Dryer}, \binits{M.}},
\bauthor{\bsnm{Fry}, \binits{C.D.}},
\bauthor{\bsnm{Sun}, \binits{W.}},
\bauthor{\bsnm{Deehr}, \binits{C.}},
\bauthor{\bsnm{Smith}, \binits{Z.}},
\bauthor{\bsnm{Akasofu}, \binits{S.-I.}},
\bauthor{\bsnm{Andrews}, \binits{M.D.}}:
\byear{2001},
\batitle{Prediction in real time of the 2000 July 14 heliospheric shock wave
  and its companions during the 'bastille' epoch}.
\bjtitle{Solar Physics}
\bvolume{204}(\bissue{1-2}),
\bfpage{267}.
\end{barticle}
\endbibitem

\bibitem[\protect\citeauthoryear{Duldig \textit{et~al.}}{1995}]{Duldig93}
\begin{bchapter}
\bauthor{\bsnm{Duldig}, \binits{M.L.}},
\bauthor{\bsnm{Cramp}, \binits{J.L.}},
\bauthor{\bsnm{Humble}, \binits{J.E.}},
\bauthor{\bsnm{Smart}, \binits{D.F.}},
\bauthor{\bsnm{Shea}, \binits{M.A.}},
\bauthor{\bsnm{Bieber}, \binits{J.W.}},
\bauthor{\bsnm{Evenson}, \binits{P.}},
\bauthor{\bsnm{Fenton}, \binits{K.B.}},
\bauthor{\bsnm{Fenton}, \binits{A.G.}},
\bauthor{\bsnm{Bendoricchio}, \binits{M.B.M.}}:
\byear{1995},
\bctitle{The ground level enhancements of 1989 September and October 22}.
In: \bbtitle{Proceedings Astronomical Society of Australia}
\bseriesno{10},
\bfpage{211}.
\end{bchapter}
\endbibitem

\bibitem[\protect\citeauthoryear{Gleeson and Axford}{1968}]{Gle68}
\begin{barticle}
\bauthor{\bsnm{Gleeson}, \binits{L.J.}},
\bauthor{\bsnm{Axford}, \binits{W.I.}}:
\byear{1968},
\batitle{Solar modulation of galactic cosmic rays}.
\bjtitle{Astrophysical Journal}
\bvolume{154},
\bfpage{1011}.
\end{barticle}
\endbibitem

\bibitem[\protect\citeauthoryear{Gopalswamy
  \textit{et~al.}}{2012}]{Gopalswamy2012}
\begin{barticle}
\bauthor{\bsnm{Gopalswamy}, \binits{N.}},
\bauthor{\bsnm{Xie}, \binits{H.}},
\bauthor{\bsnm{Yashiro}, \binits{S.}},
\bauthor{\bsnm{Akiyama}, \binits{S.}},
\bauthor{\bsnm{Mäkelä}, \binits{P.}},
\bauthor{\bsnm{Usoskin}, \binits{I.G.}}:
\byear{2012},
\batitle{Properties of ground level enhancement events and the associated solar
  eruptions during solar cycle 23}.
\bjtitle{Space Science Reviews}
\bvolume{171}(\bissue{1-4}),
\bfpage{23}.

\end{barticle}
\endbibitem

\bibitem[\protect\citeauthoryear{Hatton}{1971}]{Hat71}
\begin{bchapter}
\bauthor{\bsnm{Hatton}, \binits{C.}}:
\byear{1971},
\bctitle{The neutron monitor}.
In: \bbtitle{Progress in Elementary Particle and Cosmic-ray Physics X},
\bpublisher{North Holland Publishing Co.},
\blocation{Amsterdam}.
\bcomment{Chap. 1}.
\end{bchapter}
\endbibitem

\bibitem[\protect\citeauthoryear{Humble \textit{et~al.}}{1991}]{Humble91}
\begin{barticle}
\bauthor{\bsnm{Humble}, \binits{J.E.}},
\bauthor{\bsnm{Duldig}, \binits{M.L.}},
\bauthor{\bsnm{Smart}, \binits{D.F.}},
\bauthor{\bsnm{Shea}, \binits{M.A.}}:
\byear{1991},
\batitle{Detection of 0.5–15 GeV solar protons on 29 September 1989 at
  australian stations}.
\bjtitle{Geophysical Research Letters}
\bvolume{18}(\bissue{4}),
\bfpage{737}.
\end{barticle}
\endbibitem

\bibitem[\protect\citeauthoryear{Kallenrode, Cliver, and
  Wibberenz}{1992}]{Kallenrode1992}
\begin{barticle}
\bauthor{\bsnm{Kallenrode}, \binits{M.-B.}},
\bauthor{\bsnm{Cliver}, \binits{E.W.}},
\bauthor{\bsnm{Wibberenz}, \binits{G.}}:
\byear{1992},
\batitle{Composition and azimuthal spread of solar energetic particles from
  impulsive and gradual flares}.
\bjtitle{Astrophysical Journal}
\bvolume{391}(\bissue{1}),
\bfpage{370}.
\end{barticle}
\endbibitem

\bibitem[\protect\citeauthoryear{Klein \textit{et~al.}}{2001}]{Klein2001a}
\begin{barticle}
\bauthor{\bsnm{Klein}, \binits{K.-L.}},
\bauthor{\bsnm{Trottet}, \binits{G.}},
\bauthor{\bsnm{Lantos}, \binits{P.}},
\bauthor{\bsnm{Delaboudinière}, \binits{J.-P.}}:
\byear{2001},
\batitle{Coronal electron acceleration and relativistic proton production
  during the 14 July 2000 flare and CME}.
\bjtitle{Astronomy and Astrophysics}
\bvolume{373}(\bissue{3}),
\bfpage{1073}.
\end{barticle}
\endbibitem

\bibitem[\protect\citeauthoryear{Kudela and Usoskin}{2004}]{Kud04}
\begin{barticle}
\bauthor{\bsnm{Kudela}, \binits{K.}},
\bauthor{\bsnm{Usoskin}, \binits{I.}}:
\byear{2004},
\batitle{On magnetospheric transmissivity of cosmic rays}.
\bjtitle{Czechoslovak Journal of Physics}
\bvolume{54}(\bissue{2}),
\bfpage{239}.

\end{barticle}
\endbibitem

\bibitem[\protect\citeauthoryear{Kudela, Bu\v{c}ik, and Bobik}{2008}]{Kud08}
\begin{barticle}
\bauthor{\bsnm{Kudela}, \binits{K.}},
\bauthor{\bsnm{Bu\v{c}ik}, \binits{R.}},
\bauthor{\bsnm{Bobik}, \binits{P.}}:
\byear{2008},
\batitle{On transmissivity of low energy cosmic rays in disturbed
  magnetosphere}.
\bjtitle{Advances in Space Research}
\bvolume{42}(\bissue{7}),
\bfpage{1300}.

\end{barticle}
\endbibitem

\bibitem[\protect\citeauthoryear{Langel}{1987}]{Lan87}
\begin{bchapter}
\bauthor{\bsnm{Langel}, \binits{R.A.}}:
\byear{1987},
\bctitle{Main field in geomagnetism}.
In: \bbtitle{Geomagnetism},
\bpublisher{J.A. Jacobs Academic Press},
\blocation{London},
\bfpage{249}.
\bcomment{Chap. 1}.
\end{bchapter}
\endbibitem

\bibitem[\protect\citeauthoryear{Levenberg}{1944}]{Lev44}
\begin{barticle}
\bauthor{\bsnm{Levenberg}, \binits{K.}}:
\byear{1944},
\batitle{A method for the solution of certain non-linear problems in least
  squares}.
\bjtitle{Quarterly of Applied Mathematics}
\bvolume{2},
\bfpage{164}.
\end{barticle}
\endbibitem

\bibitem[\protect\citeauthoryear{Lockwood, Debrunner, and
  Fl{\"u}kiger}{1990}]{Lockwood1990}
\begin{barticle}
\bauthor{\bsnm{Lockwood}, \binits{J.A.}},
\bauthor{\bsnm{Debrunner}, \binits{H.}},
\bauthor{\bsnm{Fl{\"u}kiger}, \binits{E.O.}}:
\byear{1990},
\batitle{Indications for diffusive coronal shock acceleration of protons in
  selected solar cosmic ray events}.
\bjtitle{Journal of Geophysical Research: Space Physics}
\bvolume{95}(\bissue{A4}),
\bfpage{4187}.
\end{barticle}
\endbibitem

\bibitem[\protect\citeauthoryear{Marquardt}{1963}]{Mar63}
\begin{barticle}
\bauthor{\bsnm{Marquardt}, \binits{D.}}:
\byear{1963},
\batitle{An algorithm for least-squares estimation of nonlinear parameters}.
\bjtitle{SIAM Journal on Applied Mathematics}
\bvolume{11}(\bissue{2}),
\bfpage{431}.
\end{barticle}
\endbibitem

\bibitem[\protect\citeauthoryear{Mavromichalaki \textit{et~al.}}{2011}]{Mav11}
\begin{barticle}
\bauthor{\bsnm{Mavromichalaki}, \binits{H.}},
\bauthor{\bsnm{Papaioannou}, \binits{A.}},
\bauthor{\bsnm{Plainaki}, \binits{C.}},
\bauthor{\bsnm{Sarlanis}, \binits{C.}},
\bauthor{\bsnm{Souvatzoglou}, \binits{G.}},
\bauthor{\bsnm{Gerontidou}, \binits{M.}},
\bauthor{\bsnm{Papailiou}, \binits{M.}},
\bauthor{\bsnm{Eroshenko}, \binits{E.}},
\bauthor{\bsnm{Belov}, \binits{A.}},
\bauthor{\bsnm{Yanke}, \binits{V.}},
\bauthor{\bsnm{Fl{\"u}ckiger}, \binits{E.O.}},
\bauthor{\bsnm{B{\"u}tikofer}, \binits{R.}},
\bauthor{\bsnm{Parisi}, \binits{M.}},
\bauthor{\bsnm{Storini}, \binits{M.}},
\bauthor{\bsnm{Klein}, \binits{K.-L.}},
\bauthor{\bsnm{Fuller}, \binits{N.}},
\bauthor{\bsnm{Steigies}, \binits{C.T.}},
\bauthor{\bsnm{Rother}, \binits{O.M.}},
\bauthor{\bsnm{Heber}, \binits{B.}},
\bauthor{\bsnm{Wimmer-Schweingruber}, \binits{R.F.}},
\bauthor{\bsnm{Kudela}, \binits{K.}},
\bauthor{\bsnm{Strharsky}, \binits{I.}},
\bauthor{\bsnm{Langer}, \binits{R.}},
\bauthor{\bsnm{Usoskin}, \binits{I.}},
\bauthor{\bsnm{Ibragimov}, \binits{A.}},
\bauthor{\bsnm{Chilingaryan}, \binits{A.}},
\bauthor{\bsnm{Hovsepyan}, \binits{G.}},
\bauthor{\bsnm{Reymers}, \binits{A.}},
\bauthor{\bsnm{Yeghikyan}, \binits{A.}},
\bauthor{\bsnm{Kryakunova}, \binits{O.}},
\bauthor{\bsnm{Dryn}, \binits{E.}},
\bauthor{\bsnm{Nikolayevskiy}, \binits{N.}},
\bauthor{\bsnm{Dorman}, \binits{L.}},
\bauthor{\bsnm{Pustil'Nik}, \binits{L.}}:
\byear{2011},
\batitle{Applications and usage of the real-time neutron monitor database}.
\bjtitle{Advances of Space Research}
\bvolume{47},
\bfpage{2210}.

\end{barticle}
\endbibitem


\bibitem[\protect\citeauthoryear{Mironova \textit{et~al.}}{2015}]{Mironova2015}
\begin{barticle}
\bauthor{\bsnm{Mironova}, \binits{I.}},
\bauthor{\bsnm{Aplin}, \binits{K.}},
\bauthor{\bsnm{Arnold}, \binits{F.}},
\bauthor{\bsnm{Bazilevskaya}, \binits{G.}},
\bauthor{\bsnm{Harrison}, \binits{R.}},
\bauthor{\bsnm{Krivolutsky}, \binits{A.}},
\bauthor{\bsnm{Nicoll}, \binits{K.}},
\bauthor{\bsnm{Rozanov}, \binits{E.}},
\bauthor{\bsnm{Turunen}, \binits{E.}},
\bauthor{\bsnm{Usoskin}, \binits{I.}}:
\byear{2015},
\batitle{Energetic particle influence on the Earth’s
  atmosphere},
\bjtitle{Space Science Reviews}
\bvolume{194}(\bissue{1}),
\bfpage{1}.

\end{barticle}
\endbibitem


\bibitem[\protect\citeauthoryear{Mishev and Usoskin}{2013}]{Mis13a}
\begin{barticle}
\bauthor{\bsnm{Mishev}, \binits{A.}},
\bauthor{\bsnm{Usoskin}, \binits{I.}}:
\byear{2013},
\batitle{Computations of cosmic ray propagation in the Earth's atmosphere,
  towards a GLE analysis}.
\bjtitle{Journal of Physics: Conference Series}
\bvolume{409},
\bfpage{012152}.

\end{barticle}
\endbibitem

\bibitem[\protect\citeauthoryear{Mishev and Velinov}{2015a}]{Mishev2015b}
\begin{barticle}
\bauthor{\bsnm{Mishev}, \binits{A.L.}},
\bauthor{\bsnm{Velinov}, \binits{P.I.Y.}}:
\byear{2015}a,
\batitle{Time evolution of ionization effect due to cosmic rays in terrestrial
  atmosphere during GLE 70}.
\bjtitle{Journal of Atmospheric and Solar-Terrestrial Physics}
\bvolume{129},
\bfpage{78}.

\end{barticle}
\endbibitem



\bibitem[\protect\citeauthoryear{Mishev and Velinov}{2015b}]{Mishev2015c}
\begin{barticle}
\bauthor{\bsnm{Mishev}, \binits{A.L.}},
\bauthor{\bsnm{Velinov}, \binits{P.I.Y.}}:
\byear{2015}b,
\batitle{Ionization rate profiles due to solar and galactic cosmic rays during GLE 59 on Bastille day 14 July 2000}.
\bjtitle{Comptes rendus de l'Acad\'emie bulgare des Sciences}
\bvolume{68}(\bissue{3}),
\bfpage{359}.

\end{barticle}
\endbibitem


\bibitem[\protect\citeauthoryear{Mishev and Velinov}{2015c}]{Mishev2015d}
\begin{barticle}
\bauthor{\bsnm{Mishev}, \binits{A.L.}},
\bauthor{\bsnm{Velinov}, \binits{P.I.Y.}}:
\byear{2015}c,
\batitle{Determination of Medium Time Scale Ionization Effects at Various Altitudes in the Stratosphere and Troposphere During Ground Level Enhancement Due to Solar Cosmic Rays on 13.12.2006}.
\bjtitle{Comptes rendus de l'Acad\'emie bulgare des Sciences}
\bvolume{68}(\bissue{11}),
\bfpage{1425}.

\end{barticle}
\endbibitem



\bibitem[\protect\citeauthoryear{Mishev, Kocharov, and
  Usoskin}{2014}]{Mishev14c}
\begin{barticle}
\bauthor{\bsnm{Mishev}, \binits{A.L.}},
\bauthor{\bsnm{Kocharov}, \binits{L.G.}},
\bauthor{\bsnm{Usoskin}, \binits{I.G.}}:
\byear{2014},
\batitle{Analysis of the ground level enhancement on 17 May 2012 using data
  from the global neutron monitor network}.
\bjtitle{Journal of Geophysical Research}
\bvolume{119},
\bfpage{670}.

\end{barticle}
\endbibitem

\bibitem[\protect\citeauthoryear{Mishev, Usoskin, and Kovaltsov}{2013}]{Mis13b}
\begin{barticle}
\bauthor{\bsnm{Mishev}, \binits{A.}},
\bauthor{\bsnm{Usoskin}, \binits{I.}},
\bauthor{\bsnm{Kovaltsov}, \binits{G.}}:
\byear{2013},
\batitle{Neutron monitor yield function: New improved computations}.
\bjtitle{Journal of Geophysical Research}
\bvolume{118},
\bfpage{2783}.

\end{barticle}
\endbibitem

\bibitem[\protect\citeauthoryear{Moraal and McCracken}{2012}]{Moraal12}
\begin{barticle}
\bauthor{\bsnm{Moraal}, \binits{H.}},
\bauthor{\bsnm{McCracken}, \binits{K.G.}}:
\byear{2012},
\batitle{The time structure of ground level enhancements in solar cycle 23}.
\bjtitle{Space Science Reviews}
\bvolume{171}(\bissue{1-4}),
\bfpage{85}.

\end{barticle}
\endbibitem

\bibitem[\protect\citeauthoryear{More, Garbow, and Hillstrom}{1980}]{Mor80}
\begin{botherref}
\oauthor{\bsnm{More}, \binits{G.}},
\oauthor{\bsnm{Garbow}, \binits{B.S.}},
\oauthor{\bsnm{Hillstrom}, \binits{K.E.}}:
1980,
User guide for Minpack-1.
Report ANL 80-74,
Argonne National Laboratory,
Downers Grove Township, Ill., USA.
\end{botherref}
\endbibitem

\bibitem[\protect\citeauthoryear{Nevalainen, Usoskin, and Mishev}{2013}]{Nev13}
\begin{barticle}
\bauthor{\bsnm{Nevalainen}, \binits{J.}},
\bauthor{\bsnm{Usoskin}, \binits{I.}},
\bauthor{\bsnm{Mishev}, \binits{A.}}:
\byear{2013},
\batitle{Eccentric dipole approximation of the geomagnetic field: Application
  to cosmic ray computations}.
\bjtitle{Advances in Space Research}
\bvolume{52}(\bissue{1}),
\bfpage{22}.

\end{barticle}
\endbibitem

\bibitem[\protect\citeauthoryear{Perez-Peraza
  \textit{et~al.}}{2008}]{Perez-Peraza2008}
\begin{barticle}
\bauthor{\bsnm{Perez-Peraza}, \binits{J.A.}},
\bauthor{\bsnm{Vashenyuk}, \binits{E.V.}},
\bauthor{\bsnm{Gallegos-Cruz}, \binits{A.}},
\bauthor{\bsnm{Balabin}, \binits{Y.V.}},
\bauthor{\bsnm{Miroshnichenko}, \binits{L.I.}}:
\byear{2008},
\batitle{Relativistic proton production at the sun in the 20 January 2005 solar
  event}.
\bjtitle{Advances in Space Research}
\bvolume{41}(\bissue{6}),
\bfpage{947}.

\end{barticle}
\endbibitem

\bibitem[\protect\citeauthoryear{Plainaki \textit{et~al.}}{2007}]{Plainaki2007}
\begin{botherref}
\oauthor{\bsnm{Plainaki}, \binits{C.}},
\oauthor{\bsnm{Belov}, \binits{A.}},
\oauthor{\bsnm{Eroshenko}, \binits{E.}},
\oauthor{\bsnm{Mavromichalaki}, \binits{H.}},
\oauthor{\bsnm{Yanke}, \binits{V.}}:
2007,
Modeling ground level enhancements: Event of 20 January 2005.
\textit{Journal of Geophysical Research: Space Physics}
\textbf{112}(4).

\end{botherref}
\endbibitem

\bibitem[\protect\citeauthoryear{Plainaki \textit{et~al.}}{2009}]{Plainaki2009}
\begin{barticle}
\bauthor{\bsnm{Plainaki}, \binits{C.}},
\bauthor{\bsnm{Mavromichalaki}, \binits{H.}},
\bauthor{\bsnm{Belov}, \binits{A.}},
\bauthor{\bsnm{Eroshenko}, \binits{E.}},
\bauthor{\bsnm{Yanke}, \binits{V.}}:
\byear{2009},
\batitle{Modeling the solar cosmic ray event of 13 December 2006 using ground
  level neutron monitor data}.
\bjtitle{Advances in Space Research}
\bvolume{43}(\bissue{4}),
\bfpage{474}.

\end{barticle}
\endbibitem

\bibitem[\protect\citeauthoryear{Reames}{1999}]{Reames99}
\begin{barticle}
\bauthor{\bsnm{Reames}, \binits{D.V.}}:
\byear{1999},
\batitle{Particle acceleration at the sun and in the heliosphere}.
\bjtitle{Space Science Reviews}
\bvolume{90}(\bissue{3-4}),
\bfpage{413}.
\end{barticle}
\endbibitem

\bibitem[\protect\citeauthoryear{Reames}{2009a}]{Reames2009b}
\begin{barticle}
\bauthor{\bsnm{Reames}, \binits{D.V.}}:
\byear{2009}a,
\batitle{Solar energetic-particle release times in historic ground-level
  events}.
\bjtitle{Astrophysical Journal}
\bvolume{706}(\bissue{1}),
\bfpage{844}.

\end{barticle}
\endbibitem

\bibitem[\protect\citeauthoryear{Reames}{2009b}]{Reames2009a}
\begin{barticle}
\bauthor{\bsnm{Reames}, \binits{D.V.}}:
\byear{2009}b,
\batitle{Solar release times of energetic particles in ground-level events}.
\bjtitle{Astrophysical Journal}
\bvolume{693}(\bissue{1}),
\bfpage{812}.

\end{barticle}
\endbibitem

\bibitem[\protect\citeauthoryear{Shea and Smart}{1982}]{Shea82}
\begin{barticle}
\bauthor{\bsnm{Shea}, \binits{M.A.}},
\bauthor{\bsnm{Smart}, \binits{D.F.}}:
\byear{1982},
\batitle{Possible evidence for a rigidity-dependent release of relativistic
  protons from the solar corona}.
\bjtitle{Space Science Reviews}
\bvolume{32},
\bfpage{251}.
\end{barticle}
\endbibitem

\bibitem[\protect\citeauthoryear{Shea and Smart}{1990}]{Shea90}
\begin{barticle}
\bauthor{\bsnm{Shea}, \binits{M.A.}},
\bauthor{\bsnm{Smart}, \binits{D.F.}}:
\byear{1990},
\batitle{A summary of major solar proton events}.
\bjtitle{Solar Physics}
\bvolume{127},
\bfpage{297}.
\end{barticle}
\endbibitem

\bibitem[\protect\citeauthoryear{Simpson, Fonger, and Treiman}{1953}]{Sim53}
\begin{barticle}
\bauthor{\bsnm{Simpson}, \binits{J.}},
\bauthor{\bsnm{Fonger}, \binits{W.}},
\bauthor{\bsnm{Treiman}, \binits{S.}}:
\byear{1953},
\batitle{Cosmic radiation intensity-time variation and their origin. i. neutron
  intensity variation method and meteorological factors}.
\bjtitle{Physical Review}
\bvolume{90},
\bfpage{934}.
\end{barticle}
\endbibitem

\bibitem[\protect\citeauthoryear{Smart, Shea, and Fl{\"u}ckiger}{2000}]{Sma00}
\begin{barticle}
\bauthor{\bsnm{Smart}, \binits{D.F.}},
\bauthor{\bsnm{Shea}, \binits{M.A.}},
\bauthor{\bsnm{Fl{\"u}ckiger}, \binits{E.O.}}:
\byear{2000},
\batitle{Magnetospheric models and trajectory computations}.
\bjtitle{Space Science Reviews}
\bvolume{93}(\bissue{1}),
\bfpage{305}.
\end{barticle}
\endbibitem

\bibitem[\protect\citeauthoryear{Tsyganenko}{1989}]{Tsy89}
\begin{barticle}
\bauthor{\bsnm{Tsyganenko}, \binits{N.A.}}:
\byear{1989},
\batitle{A magnetospheric magnetic field model with a warped tail current
  sheet}.
\bjtitle{Planetary and Space Science}
\bvolume{37}(\bissue{1}),
\bfpage{5}.
\end{barticle}
\endbibitem

\bibitem[\protect\citeauthoryear{Tylka and Dietrich}{2009}]{Tylka09}
\begin{bchapter}
\bauthor{\bsnm{Tylka}, \binits{A.}},
\bauthor{\bsnm{Dietrich}, \binits{W.}}:
\byear{2009},
\bctitle{A new and comprehensive analysis of proton spectra in ground-level
  encahnced (GLE) solar particle}.
In: \bbtitle{Proc. of 31th ICRC Lodz, Poland, 7 -15 July 2009}.
\end{bchapter}
\endbibitem

\bibitem[\protect\citeauthoryear{Usoskin, Bazilevskaya, and
  Kovaltsov}{2011}]{Uso11a}
\begin{barticle}
\bauthor{\bsnm{Usoskin}, \binits{I.G.}},
\bauthor{\bsnm{Bazilevskaya}, \binits{G.A.}},
\bauthor{\bsnm{Kovaltsov}, \binits{G.A.}}:
\byear{2011},
\batitle{Solar modulation parameter for cosmic rays since 1936 reconstructed
  from ground-based neutron monitors and ionization chambers}.
\bjtitle{Journal of Geophysical Research}
\bvolume{116},
\bfpage{A02104}.

\end{barticle}
\endbibitem

\bibitem[\protect\citeauthoryear{Usoskin \textit{et~al.}}{2011}]{Uso11b}
\begin{barticle}
\bauthor{\bsnm{Usoskin}, \binits{I.G.}},
\bauthor{\bsnm{Kovaltsov}, \binits{G.A.}},
\bauthor{\bsnm{Mironova}, \binits{I.A.}},
\bauthor{\bsnm{Tylka}, \binits{A.J.}},
\bauthor{\bsnm{Dietrich}, \binits{W.F.}}:
\byear{2011},
\batitle{Ionization effect of solar particle GLE events in low and middle
  atmosphere}.
\bjtitle{Atmospheric Chemistry and Physics}
\bvolume{11},
\bfpage{1979}.

\end{barticle}
\endbibitem

\bibitem[\protect\citeauthoryear{Usoskin \textit{et~al.}}{2005}]{Uso05}
\begin{barticle}
\bauthor{\bsnm{Usoskin}, \binits{I.}},
\bauthor{\bsnm{Alanko-Huotari}, \binits{K.}},
\bauthor{\bsnm{Kovaltsov}, \binits{G.}},
\bauthor{\bsnm{Mursula}, \binits{K.}}:
\byear{2005},
\batitle{Heliospheric modulation of cosmic rays: Monthly reconstruction for
  1951-2004,}.
\bjtitle{Journal of Geophysical Research}
\bvolume{110}(\bissue{A12108}).

\end{barticle}
\endbibitem


\bibitem[\protect\citeauthoryear{Usoskin and
  Kovaltsov}{2006}]{Usoskin2006}
\begin{barticle}
\bauthor{\bsnm{Usoskin}, \binits{I.}},
\bauthor{\bsnm{Kovaltsov}, \binits{G.}}:
\byear{2006},
\batitle{Cosmic ray induced ionization in the atmosphere: Full
  modeling and practical applications,}.
\bjtitle{Journal of Geophysical Research}
\bvolume{111}(\bissue{D21206}).
\end{barticle}
\endbibitem



\bibitem[\protect\citeauthoryear{Usoskin \textit{et~al.}}{2015}]{Usoskin2015a}
\begin{bchapter}
\bauthor{\bsnm{Usoskin}, \binits{I.G.}},
\bauthor{\bsnm{Ibragimov}, \binits{A.}},
\bauthor{\bsnm{Shea}, \binits{M.A.}},
\bauthor{\bsnm{Smart}, \binits{D.F.}}:
\byear{2015},
\bctitle{Database of ground level enhancements (GLE) of high energy solar
  proton events}.
In: \bbtitle{Proc. of 34th ICRC Hague, Netherlands, 30 July -6 August 2015}
\bseriesno{PoS},
\bfpage{paper 54}.
\end{bchapter}
\endbibitem

\bibitem[\protect\citeauthoryear{Vashenyuk \textit{et~al.}}{2006a}]{Vas06b}
\begin{barticle}
\bauthor{\bsnm{Vashenyuk}, \binits{E.V.}},
\bauthor{\bsnm{Balabin}, \binits{Y.V.}},
\bauthor{\bsnm{Gvozdevskii}, \binits{B.B.}},
\bauthor{\bsnm{Karpov}, \binits{S.N.}}:
\byear{2006}a,
\batitle{Relativistic solar protons in the event of January 20, 2005: Model
  studies}.
\bjtitle{Geomagnetism and Aeronomy}
\bvolume{46}(\bissue{4}),
\bfpage{424}.

\end{barticle}
\endbibitem

\bibitem[\protect\citeauthoryear{Vashenyuk \textit{et~al.}}{2006b}]{Vas06}
\begin{barticle}
\bauthor{\bsnm{Vashenyuk}, \binits{E.V.}},
\bauthor{\bsnm{Balabin}, \binits{Y.V.}},
\bauthor{\bsnm{Perez-Peraza}, \binits{J.}},
\bauthor{\bsnm{Gallegos-Cruz}, \binits{A.}},
\bauthor{\bsnm{Miroshnichenko}, \binits{L.I.}}:
\byear{2006}b,
\batitle{Some features of the sources of relativistic particles at the sun in
  the solar cycles 21-23}.
\bjtitle{Advances Space Research}
\bvolume{38}(\bissue{3}),
\bfpage{411}.

\end{barticle}
\endbibitem

\bibitem[\protect\citeauthoryear{Vashenyuk \textit{et~al.}}{2008}]{Vas08}
\begin{barticle}
\bauthor{\bsnm{Vashenyuk}, \binits{E.V.}},
\bauthor{\bsnm{Balabin}, \binits{Y.V.}},
\bauthor{\bsnm{Gvozdevsky}, \binits{B.B.}},
\bauthor{\bsnm{Schur}, \binits{L.I.}}:
\byear{2008},
\batitle{Characteristics of relativistic solar cosmic rays during the event of
  December 13, 2006}.
\bjtitle{Geomagnetism and Aeronomy}
\bvolume{48}(\bissue{2}),
\bfpage{149}.

\end{barticle}
\endbibitem


\bibitem[\protect\citeauthoryear{Velinov \textit{et~al.}}{2013}]{Velinov2013}
\begin{barticle}
\bauthor{\bsnm{Velinov}, \binits{P.}},
\bauthor{\bsnm{Asenovski}, \binits{S.}},
\bauthor{\bsnm{Kudela}, \binits{K.}},
\bauthor{\bsnm{Lastovi\v{c}ka}, \binits{J.}},
\bauthor{\bsnm{Mateev}, \binits{L.}},
\bauthor{\bsnm{Mishev}, \binits{A.}},
\bauthor{\bsnm{Tonev}, \binits{P.}}:
\byear{2013},
\batitle{Impact of cosmic rays and solar energetic particles
  on the earth's ionosphere and atmosphere}.
\bjtitle{Journal of Space Weather and Space Climate}.
\bjtitle{Journal of Space Weather and Space Climate}
\bvolume{3}(\bissue{A14}).
\end{barticle}
\endbibitem

\bibitem[\protect\citeauthoryear{\v{Z}igman, Kudela, and
  Grubor}{2014}]{Zigman2014}
\begin{barticle}
\bauthor{\bsnm{\v{Z}igman}, \binits{V.}},
\bauthor{\bsnm{Kudela}, \binits{K.}},
\bauthor{\bsnm{Grubor}, \binits{D.}}:
\byear{2014},
\batitle{Response of the earth's lower ionosphere to the ground level
  enhancement event of December 13, 2006}.
\bjtitle{Advances in Space Research}
\bvolume{53}(\bissue{5}),
\bfpage{763}.

\end{barticle}

\endbibitem

\end{thebibliography}
\end{document}